\documentclass{emulateapj}
\slugcomment{{\sc Accepted to AJ:} April 23, 2012}

\usepackage{xspace}
\usepackage{amssymb}
\usepackage{graphicx}
\usepackage{booktabs}
\usepackage{hyperref}
\usepackage{txfonts}
\usepackage{hyperref}
\usepackage{graphicx, float}
\usepackage{gensymb}
\usepackage{natbib}
\usepackage{epsfig, natbib}

\newcommand\lta{\mathrel{\hbox{\raise 0.6 ex \hbox{$<$}\kern
                   -1.8 ex\lower .5 ex\hbox{$\sim$}}}}
\newcommand\gta{\mathrel{\hbox{\raise 0.6 ex \hbox{$>$}\kern
                   -1.7 ex\lower .5 ex\hbox{$\sim$}}}}

\shorttitle{A New Reddening Law for M4}
\shortauthors{Hendricks et al.}

\begin{document}

\title
{A New Reddening Law for M4} 
\author{Benjamin Hendricks\altaffilmark{1}, Peter B. Stetson\altaffilmark{2}, Don A. VandenBerg\altaffilmark{1}, Massimo Dall'Ora\altaffilmark{3}}

\affil{$^1$Department of Physics \& Astronomy, University of Victoria, P.O. Box 3055, Victoria, BC V8W 3P6}
\affil{$^2$National Research Council Canada, 5071 West Saanich Road, Victoria, BC V9E 2E7}
\affil{$^3$INAF-Osservatorio Astronomico di Capodimonte, Via Moiarello 16, 80131, Naples, Italy}

\begin{abstract}
We have used a combination of broad-band near-infrared and optical Johnson-Cousins photometry to study the dust properties in the line of sight to the Galactic globular cluster M4. We have investigated the reddening effects in terms of absolute strength and variation across the cluster field, as well as the shape of the reddening law defined by the type of dust. All three aspects had been poorly defined for this system and, consequently, there has been controversy about the absolute distance to this globular cluster, which is closest to the Sun.
Here, we determine the ratio of absolute to selective extinction ($R_V$) in the line of sight towards M4, which is known to be a useful indicator for the type of dust and therefore characterizes the applicable reddening law. Our method is independent of age assumptions and appears to be significantly more precise and accurate than previous approaches.
We obtain ${A_V / E(B-V)}=3.76\pm0.07$ (random error) for the dust in the line of sight to M4 for our set of filters. That corresponds to a dust-type parameter $R_V=3.62\pm0.07$ in the \citet{Cardelli89} reddening law. With this value, the distance to M4 is found to be $1.80\pm0.05\,\mbox{kpc}$, corresponding to a true distance modulus of $(m-M)_{0}=11.28\pm0.06$ (random error). 
A reddening map for M4 has been created, which reveals a spatial differential reddening of $\delta E(B-V)\geq0.2\,\mbox{mag}$ across the field within $10^{\prime}$ around the cluster centre; this is about 50\% of the total mean reddening, which we have determined to be $E(B-V)=0.37\pm0.01$.
In order to provide accurate zero points for the extinction coefficients of our photometric filters, we investigated the impact of stellar parameters such as temperature, surface gravity and metallicity on the extinction properties and the necessary corrections in different bandpasses. 
Using both synthetic ATLAS9 spectra and observed spectral energy distributions, we found similar sized effects for the range of temperature and surface gravity typical of globular cluster stars: each causes a change of about $3\%$ in the necessary correction factor for each filter combination. Interestingly, variations in the metallicity cause effects of the same order when the assumed value is changed from the solar metallicity ($\mbox{[Fe/H]}=0.0$) to $\mbox{[Fe/H]=--2.5}$. 
The systematic differences between the reddening corrections for a typical main-sequence turnoff star in a metal poor globular cluster and a Vega-like star are even stronger ($\sim5\%$).
We compared the results from synthetic spectra to those obtained with observed spectral energy distributions and found significant differences for temperatures lower than $\sim5\,000\,\mbox{K}$. We have attributed these discrepancies to the inadequate treatment of some molecular bands in the $B$ filter within the ATLAS9 models. Fortunately, these differences do not affect the principal astrophysical conclusions in this study, which are based on stars hotter than 5\,000\,K.
From our calculations, we provide extinction zero points for Johnson-Cousins and 2MASS filters, spanning a wide range of stellar parameters and dust types. These extinction tables are suited for accurate, object-specific extinction corrections. 
\end{abstract}

\keywords{globular clusters: individual (M4, NGC~6723) --- ISM: dust, extinction --- Techniques: photometric}

\section{Introduction}
\label{chapter_introduction}
Globular cluster systems rank among the oldest objects known in our Galaxy. Their ages set a lower limit for the age of the Universe. Unfortunately, a precise measurement of the absolute age of a globular cluster (GC) is strongly dependent on the knowledge of its precise distance and reddening (\citealt{Gratton03}; \citealt{Bolte95}).

M4 is a peculiar cluster in terms of interstellar reddening. It has a surprisingly large total amount of reddening ($E(B-V)=0.35$, \citealt{Harris10}) for its small distance ($\sim 2\,\mbox{kpc}$), which is due to its location in the Galactic plane, behind the Sco-Oph cloud complex. Moreover, the cluster suffers from a significant amount of spatially differential reddening (\citealt{Cudworth90}; \citealt{Drake94}; and \citealt{Ivans99}) which has been taken into account in some recent studies of M4 (e.g. \citealt{Marino08}, \citealt{Mucciarelli11}). In the literature, approximations for peak-to-peak differences within a distance to the cluster centre of at least $10^\prime$ range from $\delta E(B-V) \geq 0.05$ in \citet{Cudworth90} to $\delta E(B-V)=0.25$ in \citet{Mucciarelli11}.

The extinction law that is used to correct for interstellar reddening can be different for each line of sight, depending on the type of dust located between the observer and the object, 
where the total to selective extinction $R_V\equiv A_{V}/E(B-V)$ is known to be a good index of the extinction properties of a particular dust type (\citealt{Cardelli88}, \citealt{Mathis90}).
However, in many studies of star clusters a standard reddening law is applied, assuming dust characteristics typical of the diffuse interstellar medium (ISM) as the main origin of interstellar extinction: a reddening law with a value of $R_V$ close to 3.1 is typically adopted (e.g. \citealt{Sneden78}, \citealt{McCall04}). However, since $R_V$ defines both the shape of the reddening law and the ratio between extinction and reddening, a reconsideration of this standard assumption is appropriate when determining the absolute distance of a reddened stellar system. 

For M4, there are several hints in former studies where the authors suggest an abnormal dust type, or at least remark on certain discrepancies that arise when using a standard assumption of $R_{V}=3.1$. 
For example, \citet{Ivans99} compare their spectroscopically derived temperatures of bright red giant branch (RGB) stars to the temperatures derived from photometric indices and found $R_V=3.4 \pm 0.4$. \citet{Dixon93} propose an $R_V$ ``closer to 4 then to 3" by evaluating the relative location of RGB sequences from M4 compared to M3, M13 and M92 using data from \citet{Frogel83}. Interestingly, the only study that does not rely on the calibration of photometric magnitudes reaches a similar result: \citet{Peterson95} derived a geometric distance for M4 from proper motion and radial velocity measurements and find a value significantly smaller than the accepted photometric distance, a result which is only in agreement with canonical horizontal branch (HB) magnitude measurements if a reddening law with $R_V \approx 4$ is assumed. In addition to direct measurements of $R_V$ using cluster members, there are several indirect measurements using individual stars in the Sco-Oph dust cloud complex with a line of sight close to M4, which yield values of $R_V$ around 4 (see e.g., \citealt{Clayton88}, \citealt{Vrba93}).

Indeed, some studies of M4's colour-magnitude diagram (CMD) already use a non-standard, higher value of $R_V$. \citet{Richer97} (and later \citealt{Richer04}) were among the first to use a value of $R_V=3.8$ for their distance estimate when determining the age of M4 from the white dwarf cooling sequence. However, this choice is solely based on the vague statements of \citet{Peterson95}, and \citet{Vrba93} who only estimate the value to be ``around 4." Later, \citet{Hansen04} and \citet{Bedin09} follow the argumentation of Richer and additionally use the results from \citet{Clayton88}, who found a value of $R_V=3.8$ for a star only one degree away from the line of sight of M4 to justify their choice. However, \citet{Mathis90} warns in his review about the properties of interstellar dust that ``it is not possible to estimate $R_V$ quantitatively from the environment of a line of sight," since significant variations in the properties of dust can occur even on these small angular scales (see also \citealt{Vrba85}). Moreover, this value for $R_V$ is based on the measurement of only one star, using obsolete stellar models to estimate its atmospheric parameters.

M4 is the closest GC to the sun. Its closeness and low degree of crowding make the stellar population accessible to especially deep photometric and spectroscopic analyses. For example, M4 hosts one of the largest populations of identified white dwarfs (WDs), making it attractive for absolute age determinations and an excellent laboratory for testing stellar evolutionary theory (e.g. \citealt{Richer97}). It has a significant population of RR~Lyrae stars and it shows a bimodal HB, with well populated blue and red parts, making it an interesting object to study the so called ``Second Parameter Problem" (see e.g. the review by \citealt{Catelan_review_09} or some recent work by \citealt{Marino11}).

There is clearly a need for a detailed investigation of the absolute reddening and the specific reddening law for M4, providing more precise information about its dust type and the effects of spatial differential reddening across the cluster face. A better determination of the absolute photometric properties of M4's stars---and consequently its distance---will increase the cluster's status as a most attractive target for astrophysical investigations.

In this study we investigate the type of dust in the line of sight to M4, as characterized by the appropriate value of $R_V$ assuming the validity of the general reddening law given in \citet{Cardelli89}. We achieve high quantitative precision by using a combination of near infrared (NIR) $J$ and $K_s$ and optical Johnson-Cousins $UBV(RI)_c$ photometry for the cluster M4, supplemented with equivalent photometry for the cluster NGC 6723 with which we verify our models in unreddened conditions. We use the newest set of Victoria-Regina isochrones (\citealt{VandenBerg12}) to determine colour excesses in filter combinations that are sensitive to the type of dust and consequently to the shape of the reddening law. To increase the precision of parameters such as distance, age and the absolute locus of the observed fiducial sequence, we map the variations in the reddening across the field of M4 that has been surveyed and then correct for those differential effects. 

Since extinction is a function of wavelength, the accurate extinction properties of our broadband filters depend upon the distribution of observed stellar flux within the passband. To minimize systematic errors in our extinction corrections and to the determined value of $R_V$, we investigate differential changes in the extinction properties of optical and NIR filters and filter combinations appropriate for stars with different intrinsic spectral-energy distributions as determined by their temperature, surface gravity and metallicity. Specifically, we consider whether a single reddening law is sufficient to determine cluster parameters from photometry, or whether a star-by-star correction is required to avoid significant systematic errors. These results are then used to define an object-specific reddening law for M4 where the correction zero points are tailored for the atmospheric and chemical parameters of this cluster instead of assuming a Vega-like star, as is the case for most literature values. To test the consistency of the results obtained with synthetic and observed stellar fluxes we compare synthetic spectra from the ATLAS9 library (\citealt{Castelli03}) to observational databases from \citet{Pickles98} and \citet{Sanchez06}. The outcome of such a comparison not only constrains the validity of our results, but also reveals possible weak spots in the atmospheric models.

In \S\ref{chapter_datareduction}, we summarize the data and the reduction process together with the criteria for choosing our high-quality photometric sample. Our procedure for generating a reddening map for M4 and the spatial differential reddening corrections is explained in \S\ref{chapter_diffred}, while in \S\ref{chapter_reddeninglaw}, we investigate the effects of stellar temperature, surface gravity, metallicity and extinction on the shape of the reddening law and calculate an object-specific law tailored for M4. Finally, in \S\ref{chapter_dusttype} we characterize the dust type toward M4, derive the cluster's absolute distance, and discuss random and systematic errors in detail. A summary is given in \S\ref{chapter_summary}. 
Appendices provide fiducial sequences for M4 and NGC 6723, and explicit extinction tables for a wide range of $T_{\mathrm{eff}}$, [Fe/H] and $R_V$.

\section{Data and Data Reduction}
\label{chapter_datareduction}
\subsection{Instrumental Photometry}

For this work, we use near infrared $J$ and $K_s$\footnote{Throughout the remainder of this paper $K$ and $K_s$ are used interchangeably where both refer to the 2MASS bandpass.} ground-based photometry for the globular clusters M4 (NGC 6121) and NGC 6723. 
The images for M4 were taken in 2002 with the large and small field of SOFI, the infrared imaging camera on the New Technology Telescope of ESO on Cerro La Silla in Chile as part of the program 69.D-0604(A) (PI: Zoccali) and cover a field of $\sim 13^{\prime} \times 13^{\prime}$.
For NGC 6723 the same telescope and instrumentation were used in 2005 as part of the program 075.D-0372 (PI: Ferraro).
The pre-processing of the CCD images for both clusters included bad pixel masking, bias and dark frame subtraction, and flat-field corrections, as well as sky subtraction.

We use the data reduction packages DAOPHOT II (\citealt{Stetson87}; \citealt{Stetson88}) and ALLFRAME (\citealt{Stetson94}) and related routines to obtain the instrumental photometry for each cluster and transform it to the 2MASS photometric system (\citealt{Skrutskie06}).
Since the data reduction process for our two clusters is very similar, we will describe this work in detail only for M4 as an example. 

In total, 50 frames in each filter were taken with a main concentration on the central region of the cluster, which was observed 10 times, and one region outside the core, which was observed 20 times. The rest of the cluster is covered by only one exposure in each filter with slight overlap. 

For the profile-fitting photometry we use a point spread function (PSF) which varies quadratically with the position in the frame because the SOFI large-field camera was not properly aligned, so that the images are radially elongated in the left part of each frame. This distortion affects a vertical strip of about 150 pixels and smoothly disappears toward the centre of the frame.
Later, we correct the residual spatial inhomogeneities to first order by applying a differential reddening correction (see section \ref{chapter_diffred}).
The PSF is calculated using a sample of bright, isolated program stars, uniformly distributed over the individual frames. Although we have written a pipeline to process all frames through DAOPHOT and ALLFRAME automatically, we pick out the PSF stars for each frame by eye to assure their quality and rule out blending. The same stars later serve as local standards to calibrate our instrumental photometry to the 2MASS standard system. 

We use DAOMATCH and DAOMASTER to find positional transformation equations between the frames, accounting for differences in scale and rotation as well as transverse offsets. Those equations are then used to match up stars between different frames in order to get a complete star list for our catalog. 

To calibrate the frames among themselves, DAOGROW (\citealt{Stetson90}) is used to obtain the total integrated instrumental magnitudes (i.e., the total photon count) for our standard stars by multiple aperture photometry. These total magnitudes are then calibrated by direct comparison to the published 2MASS magnitudes for the same stars, as described in the next section.

\subsection{Calibration to the 2MASS Standard System} 
Our instrumental photometry for M4 must be calibrated to the 2MASS standard system: for accurate comparisons of our observed data to stellar evolutionary models, it is especially important that systematic deviations between instrumental and standard magnitudes are small.

The same hand-selected, bright and isolated program stars used to determine the PSF above now serve as local standards for our photometry. 
As an additional criterion, we only use stars below the 1\% linearity limit of the detector ($\leq 13\,000\,\mbox{ADU}$), and only those that could be cross-identified with a 2MASS standard star with a coordinate deviation less than $0.3^{\prime\prime}$. For M4, these comprise about 360 stars covering the whole observation field. Since they are distributed homogeneously over the different frames, we are able to detect and compensate for photometric differences between different exposures as well as spatial inhomogeneities within individual fields. 

\begin{figure}[t]
\begin{center}
\includegraphics[width=0.5\textwidth]{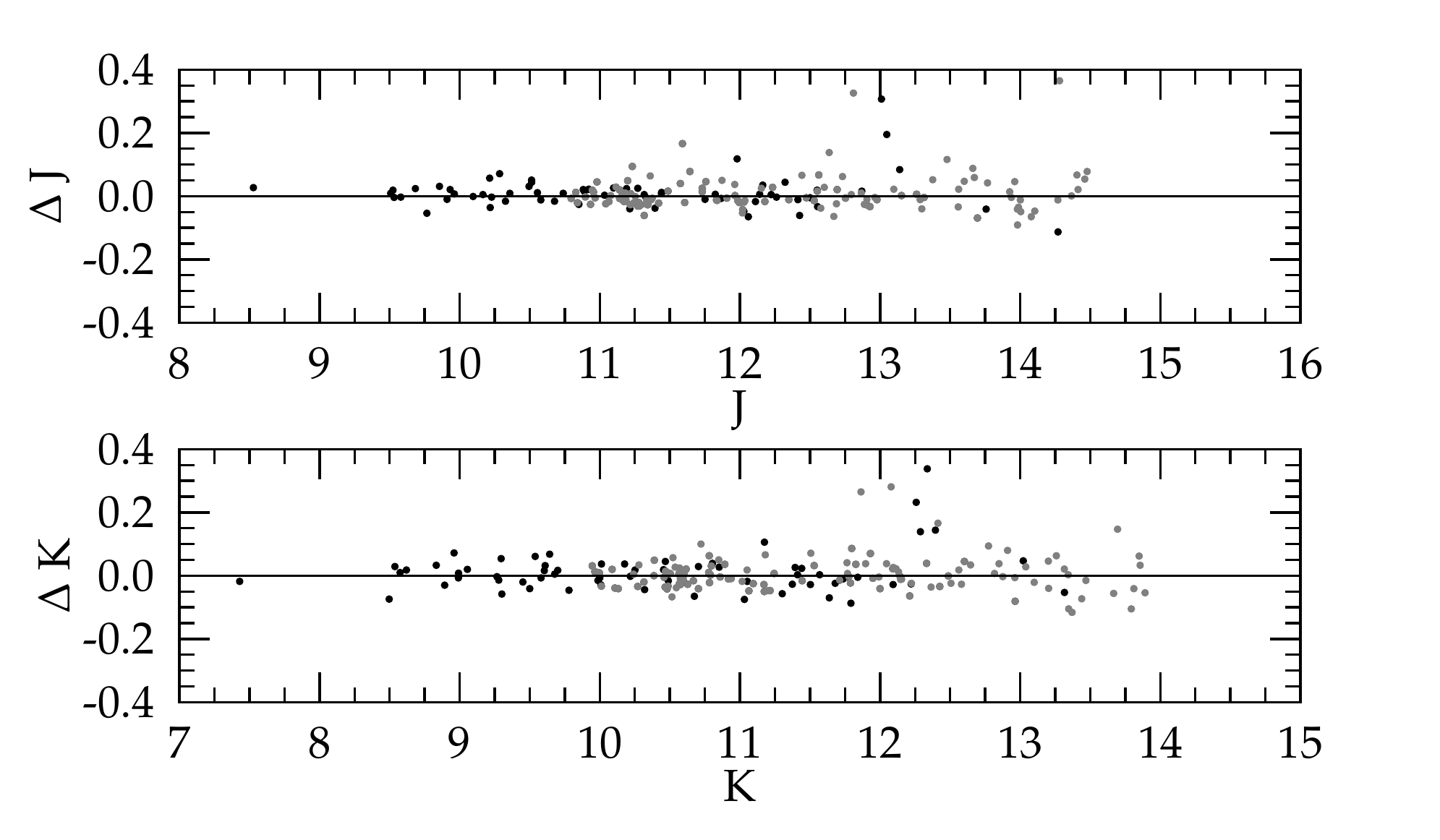}
\caption[Photometric differences between our photometry for M4 and 2MASS standard stars as a function of magnitude.]{Photometric differences between our photometry for M4 and 2MASS standard stars where $\Delta J = J-J_{2MASS}$ and $\Delta K = K-K_{2MASS}$.
Only stars are shown that could be cross-identified with a 2MASS star to within $0.3^{\prime\prime}$. The solid black line indicates a theoretical zero offset. Stars used as local standards are viewed in black, grey dots indicate all other identified 2MASS stars from our final high-quality sample (see following paragraphs).}
\label{n6121_consistency1}
\end{center}
\end{figure}

\begin{figure}[t]
\begin{center}
\includegraphics[width=0.5\textwidth]{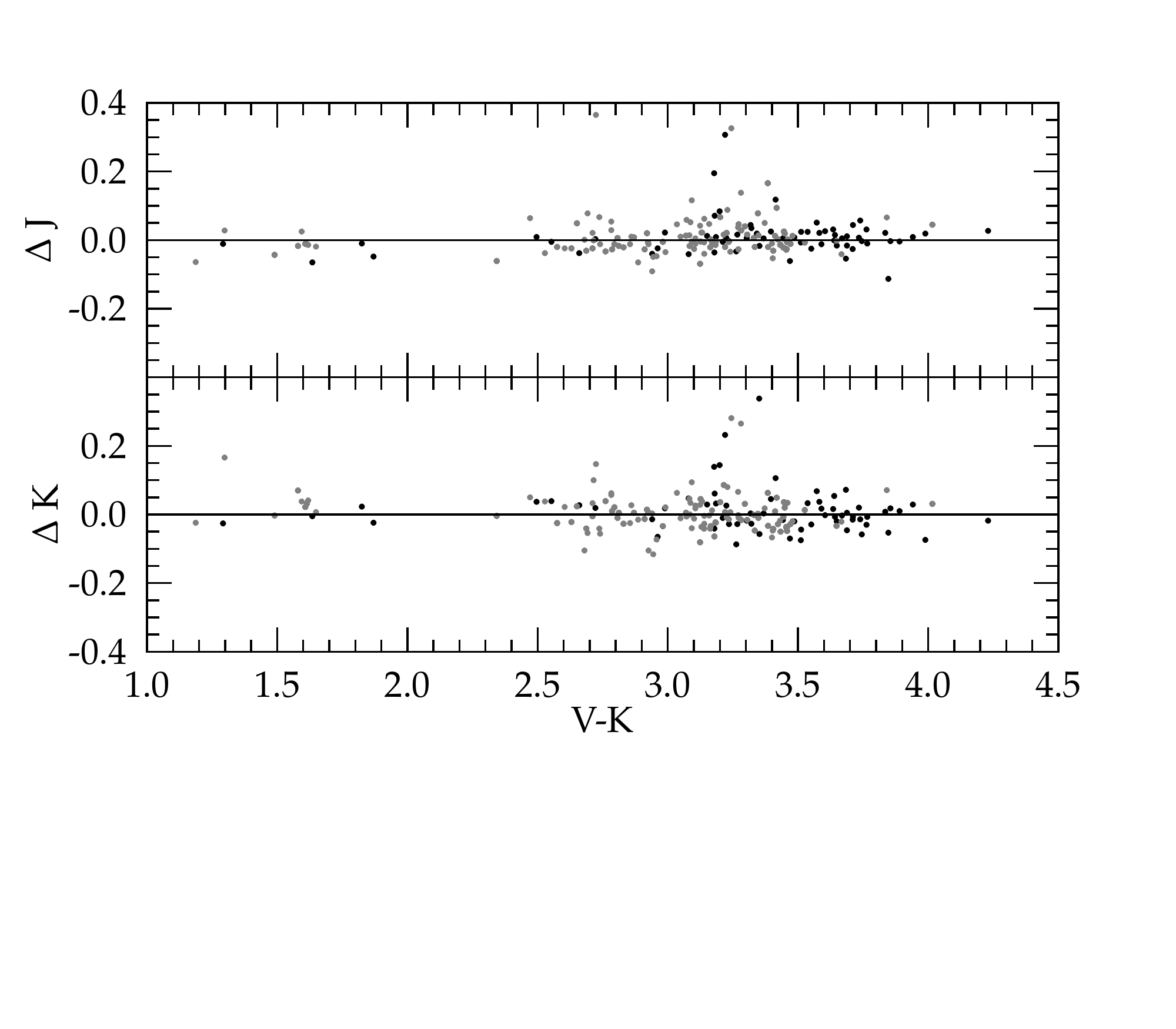}
\caption[Photometric differences between our photometry for M4 and 2MASS standard stars as a function of colour.]{As in the previous plot, except that photometric differences are plotted as a function of $V-K$ to examine whether there are any trends with temperature.}
\label{n6121_consistency2}
\end{center}
\end{figure}

DAOMASTER is then used to match up 2MASS stars with corresponding local standards in our photometry.
The equation that we use to transform the instrumental magnitudes ($m$) to the standard system ($M$) is of the general form
\begin{equation}
m_1=M_1 + z_c + a_c  X + c_c (m_1-m_2)
\end{equation}
and includes a zero-point correction ($z_c$), a term for atmospheric extinction ($a_c$; with airmass $X$), and a colour-sensitive term ($c_c$) to account for differences in filter characteristics between the two systems. For our $J$ photometry we find $z_c=1.956\pm0.004$, $a_c=0.016\pm0.034$, and $c_c=-0.092\pm0.005$. For the $K$ filter, $z_c=2.641\pm0.006$, $a_c=0.126\pm0.046$, and $c_c=0.008\pm0.006$.
The correction coefficients for atmospheric extinction ($a_c$) are a little low for $J$ and quiet high for $K$, but since the variation in airmass within our individual frames is only 0.028 for both NIR bandpasses, the impact on our photometry compared to ``standard'' assumptions of $a_c \approx 0.04$ for each filter would only be in the order of $\sim0.001$~mag and therefore negligible for our results.
Once the transformation equations to the standard system are defined for each filter with the standard star sample, they are applied to all of the stars in the cluster.

A comparison between our photometry and the 2MASS standard system is shown in Figures~\ref{n6121_consistency1} and \ref{n6121_consistency2} where the deviation between the two systems is plotted as a function of magnitude and colour. We find neither a significant zero-point offset between the two systems nor a clear trend with colour or magnitude, which would arise from insufficient correction of the different filter characteristics.

\subsection{Optical Photometry}
The optical Johnson-Cousins photometry for M4 is a compilation of 4\,794, primarily archival, images from 48 independent nights distributed among 11 different observing runs. The images were obtained between 1994 and 2007 with several different ground-based telescopes ranging from 0.9\,m to 3.6\,m, and were analyzed and calibrated as described in \citet{Stetson00} and \citet{Stetson05}.
In total, the optical database covers a field of $25^\prime \times 25^\prime$ around the cluster centre, which is about four times the area that we cover with our NIR photometry. The database consists of $\sim81\,000$ stars and the photometry reaches down to $B \sim25\,\mbox{mag}$, $V \sim23.5\,\mbox{mag}$, and $I\sim21\,\mbox{mag}$.
We use DAOMASTER to match up stars between the two photometric sets, where only stars are considered for which the coordinate deviation between the optical and the NIR database is smaller than $0.3^{\prime\prime}$.

\subsection{Selection of a High-Quality Sample}
The initial output of the photometry packages DAOPHOT and ALLFRAME yields a total of $\sim21\,000$ stars with photometric information in both $J$ and $K_s$ and in at least two of the optical bands. This sample still includes stars with high photometric errors, field stars in the line of sight to the actual GC system, and variable stars. Since we are more interested in photometric precision than in completeness, a high-quality subsample (hereafter: HQ-sample) was selected taking into consideration the aspects mentioned above.

To select only stars with the best photometric standard errors, we have determined the 20-th percentile of standard errors in magnitude bins of $\sim1$~mag
straight line segments were then interpolated between these points to define a rejection criterion as a function of magnitude.
This has been done for each filter individually to account for the different characteristics of each bandpass and the different observing conditions for each dataset. 
It is important to note that we do not automatically choose a lower boundary for brighter stars. A less strict boundary for HB stars has been applied so as to obtain a larger, more complete sample in this evolutionary stage and thereby to better to define the observed zero-age horizontal branch (ZAHB) luminosity, which is an important distance and age indicator for the cluster. Furthermore, some pixels in the brightest stars occasionally exceed the saturation limit of the detector, which increases their observed photometric standard error. To produce a clear RGB sequence extending as far as possible towards brighter magnitudes, we applied a less strict photometric rejection limit for these stars, as well. 

\begin{figure}[t]
\begin{center}
\includegraphics[width=0.5\textwidth]{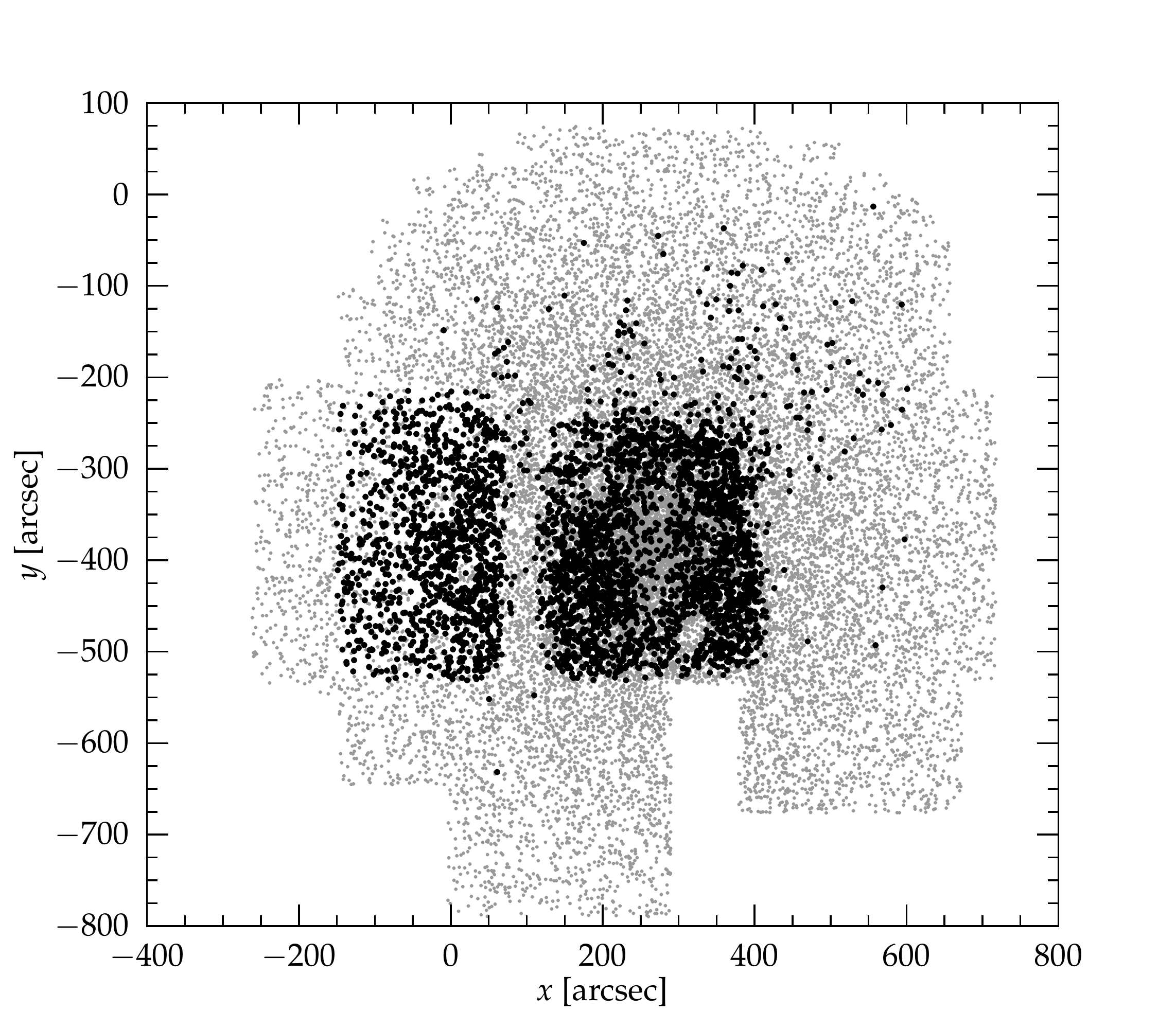}
\caption[Location of stars from the selected HQ-sample in the field of M4.]{Location of stars from the HQ-sample in the field of M4: The final selection (black) is not a homogeneous distribution over the whole cluster field (grey), but shows a preference for stars lying in areas with the most exposures.}
\label{hq_sample}
\end{center}
\end{figure}

Since the distribution of frames over the cluster field has an important impact on the final photometric standard error of a star---a large number of frames per observing field decreases the photometric error of stars therein---the final HQ-sample is not distributed homogeneously over the field, but favours areas that have been observed most frequently (see Figure~\ref{hq_sample}).

A significant fraction of field stars can be a problem in the analysis of GC CMDs, since they are not located on the evolutionary sequence of the population. Therefore, they can smear the actual sequence and, in the worst case, lead to misinterpretations of observed features. By investigating our CMDs of the whole cluster sample and comparing the total number of stars to those which clearly do not follow the common cluster sequence, it is clear that the fraction of field stars lying in the line of sight to the object is very small for the field of view of our observations. From this ratio, we infer the fraction of field stars to be $\leq 1\%$. Therefore we assume that field stars do not affect our photometric analysis. 

\begin{figure}[t]
\begin{center}
\includegraphics[width=0.5\textwidth]{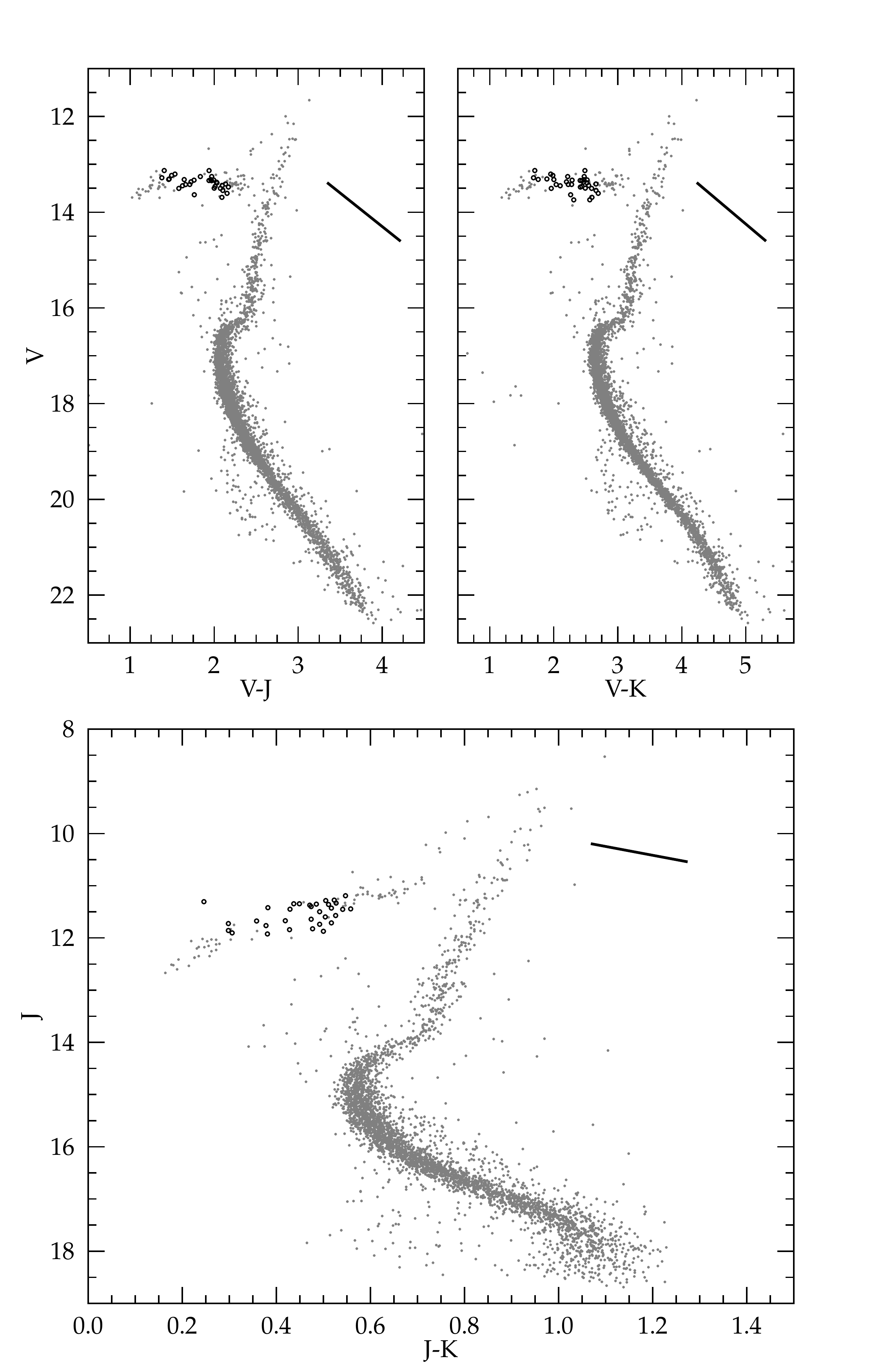}
\caption[M4 $\mathrm{(V-J)}$, $\mathrm{(V-K)}$, and $\mathrm{(J-K)} CMDs.]{Different CMDs containing the HQ-sample of M4. Black circles indicate the identified RR~Lyrae variable stars in the sample. In the bottom plot, the MS ``knee,'' which is seen only in the NIR, is just visible at $J$ $\sim 17.5\ \mbox{mag}$. The apparent larger spread in colour showed by the NIR filter combination is only caused by the smaller colour range covered by this CMD. The black bars indicate the length and direction of the reddening vector for each filter combination assuming $E(B-V)=0.37$ and a standard reddening law with $R_V=3.1$.}
\label{HQ CMDs}
\end{center}
\end{figure}

RR~Lyrae stars have been identified by their coordinates from the most recently updated database of C. Clement (\citealt{Clement10}; 2011 edition). Only cross-identifications with a coordinate deviation of less than $1^{\prime\prime}$ and only detections lying $\sim \pm 1\ \mbox{mag}$ around the ZAHB $V$-band magnitude have been considered to be true RR~Lyrae identifications. By these criteria, 35 RR~Lyrae stars were identified in our sample, which are plotted in Figure~\ref{HQ CMDs} as black circles. Except for a few outliers, these stars fall very well in the predicted instability strip. For all RR~Lyrae stars, we adopt the ALLFRAME magnitudes from our photometry and not the mean magnitudes.

In our final HQ-sample, the photometry reaches to well below the main sequence turn-off (MSTO) to a magnitude of $\sim 22.5$ in $V$ and $\sim 18.5$ in J. Notably, it reaches below the MS knee visible only in NIR photometry which is an interesting feature for the determination of the distances, and hence ages, of globular clusters (see \citealt{Bono10}). At the luminous end, the photometry is limited by the linearity limit of the NIR data which lies at $\sim 11.5\ \mbox{mag}$ in $V$ and $\sim 9.0\ \mbox{mag}$ for the $J$ filter.

\section{Differential Reddening}
\label{chapter_diffred}

In our photometry of M4, both the HB and the RGB sequences are poorly defined (seen as a large scatter in magnitude on the HB and a large scatter in colour along the RGB) compared to the well defined subgiant branch (SGB) and MS. Additionally, in CMDs with a combination of optical and NIR filters, the colour scatter of the MS increases with luminosity towards the TO. Considering only the photometric uncertainty, one would expect the result to be the opposite. Furthermore, the scatter on the HB is much stronger in $V$ than it is in the $J$-band, suggesting that those effects are mainly caused by differential reddening instead of photometric errors. Generally, differential reddening induces the stars to scatter much more around the actual sequence than would be expected from their photometric uncertainty. This additional scatter, however, is not uniformly evident throughout the GC fiducial sequence. In fact, it depends on the angle between the sequence and the reddening vector. Regions of the CMD where this angle is large show a larger scatter than regions where the sequence and the reddening vector are almost parallel. The direction of the reddening vector depends upon the actual filters that are used and the extinction relation between these filters, defined by the reddening law. In Figure~\ref{HQ CMDs} we show the reddening vector for each individual filter combination using a standard reddening law with $R_V=3.1$ for purpose of illustration. Note, that for M4 we expect a slightly different orientation of the reddening vector. 

\subsection{Correction Procedure}
\label{diffred_correctionprocedure}

The general idea for determining and correcting spatially differential reddening across the face of the cluster is to calculate for a sample of reference stars the distance from a fiducial sequence in a CMD as measured along the reddening vector. In theory, this displacement depends only upon the individual reddening of the star and is therefore directly correlated with its spatial location in the cluster field. Now, the local reddening value defined by these reference stars can be provisionally applied to all program stars in their immediate neighbourhood (see e.g., \citealt{Piotto99}, \citealt{Sarajedini07}). Such a correction for spatially differential effects in the photometry potentially includes a first-order correction for poor flat-fielding or spatially varying PSFs within different frames as well, in case they were not modelled perfectly during the data reduction process (as is likely the case for our photometry).

Our strategy for correcting differential reddening effects in M4 is a ``method of closest neighbours," where the reddening displacement for each program star is determined as the median value among the closest 10 to 30 neighbours for which the residuals have been determined. With this approach, we are able to assign a reddening value on a star-by-star basis, allowing for any arbitrary dust geometry. The resolution is automatically adjusted to the number density in a given area, determined by the neighbour with the largest distance which is used. Obviously, this method smooths over, and is therefore insensitive to, reddening variations on any angular scale smaller than 3--6 times the typical angular separation between stars.
In general, we adopt the method described by \citet{Milone11} and implement only a few small changes. 

In order to obtain the maximal discrimination between the effects of differential reddening and those due to photometric error, we use a $V$~vs.~$V-K$ CMD, which---besides the large dynamic range of this colour compared to its uncertainty---also provides another important advantage over other filter combinations. In general, the direction of the reddening vector in the CMD depends upon the value of $R_V$ that describes the reddening law. This filter combination, however, is highly insensitive to the reddening law due to the fact that the extinction in $K_s$ barely changes with the dust type. Therefore the effect on both axes is dominated by the change in $V$ and almost cancels out, with the result that we are able to perform the reddening correction without knowing $R_V$ precisely at this point. To define the angle of the reddening vector, we use ${{A_{K}}/{A_{V}}}=0.123$, assuming $R_V\approx 3.6$, but a standard assumption of 3.1 would only change the angle of the vector by less then 0.5 degree.
 
In addition to a reference star sample on the main sequence, we select a second sample along the RGB which is smaller in number but shows stronger sensitivity to spatial differential reddening because it is more nearly orthogonal to the reddening vector. 

Since our data have low photometric uncertainty only for specific regions in the field of M4, we exclude only stars with especially large uncertainties to be able to cover the majority of the surface area (see Table~\ref{reference criteria}).

\begin{table}[h]
\centering
\caption{Selection criteria for reference stars.}
\begin{tabular}{ccccc}
\hline%
\hline
	Area	 		& 	$V$		&	$\sigma (V-K)$& \# Stars	\\\hline
  MS		   		& 16.7 - 18.3    	& $\leq0.04$& 4060		\\
  RGB    			& 13.6 - 15.7	& $\leq0.03$& 444		\\\hline
\end{tabular}
\label{reference criteria}
\end{table}
 
To obtain the best angular resolution in the dense central regions and simultaneously be able to define differential reddening values for stars in sparse outer regions, we apply four different levels of significance: 
In a first step, the 30 closest neighbours to each star are selected. In case not all of them fall within a maximum radius of $R_{max}=40^{\prime\prime}$, the process is started again, this time searching only for the closest 25 neighbours. If necessary, the iteration is repeated for 20, 15, and finally 10 closest neighbours until all of them are located within $R_{max}$. 
With this approach, we assure the actual astrometric proximity of the closest neighbours to each of the selected program stars by accepting a poorer standard error of the mean reddening for values derived in sparse outer regions where fewer neighbours can be found in the local environment. 
Any stars for which fewer than 10 neighbours are located within the critical radius are not assigned a differential reddening value.

\subsection{Results}
\label{diffred_results}

\begin{figure}[t]
\begin{center}
\includegraphics[width=0.5\textwidth]{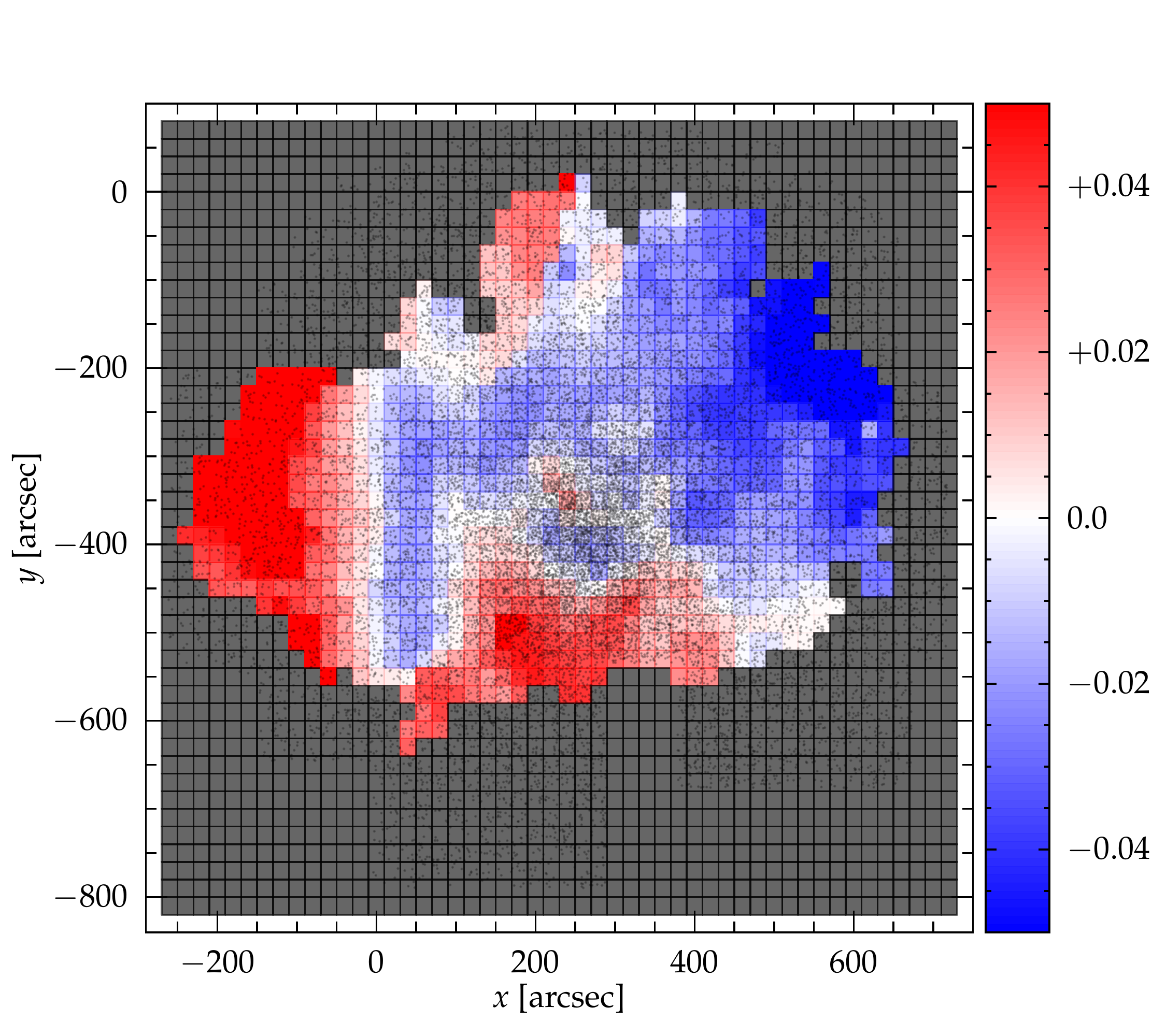}
\caption[Reddening Map of M4.]{Reddening map of M4. Blue areas indicate a lower reddening compared to the mean value and red areas indicate a higher reddening. Some differential features are presumably caused by PSF variations within observation frames and are not due to differential reddening (see the text) The coordinate system is normalized to an arbitrary zero point at $\mbox{RA}=\mathrm{16^{h}23^{m}15.12^{s}}$, $\mbox{Dec}=\mbox{26\degree}25^{\prime}15.4^{\prime\prime}$. The cluster centre is located at $(x,y)\approx(220,-400)$.}
\label{reddening map}
\end{center}
\end{figure}

To visualize the spatially differential reddening across the cluster face of M4, we bin the surface in square cells of $20^{\prime\prime} \times 20^{\prime\prime}$ and calculate the median reddening residual for all stars that fall in this coordinate range, and for which a reddening value has been assigned in the previous steps. Here, blue cells indicate regions with a reddening below the overall mean and red cells regions with values above, while the colour intensity increases with increasing absolute value. Cells containing only undetermined stars are shaded dark grey. Cells saturate when the differential reddening value becomes larger than $|\Delta E(B-V)|=0.05$. The resulting reddening map is shown in Figure~\ref{reddening map}. 

We are able to determine spatially differential effects for a total area of about $10^{\prime} \times 10^{\prime}$ around the centre of M4. We estimate the size of the differential reddening by evaluating the median values for stars within cells of $40^{\prime\prime} \times 40^{\prime\prime}$ and find a total range of $\approx 0.2\,\mbox{mag}$ between the lowest and highest reddening values within the area analyzed. That is more than half of the total mean reddening of M4 which we determine to be $E(B-V)=0.37$ (see section \ref{dusttype_results}).

\begin{figure}[t]
\begin{center}
\includegraphics[width=0.5\textwidth]{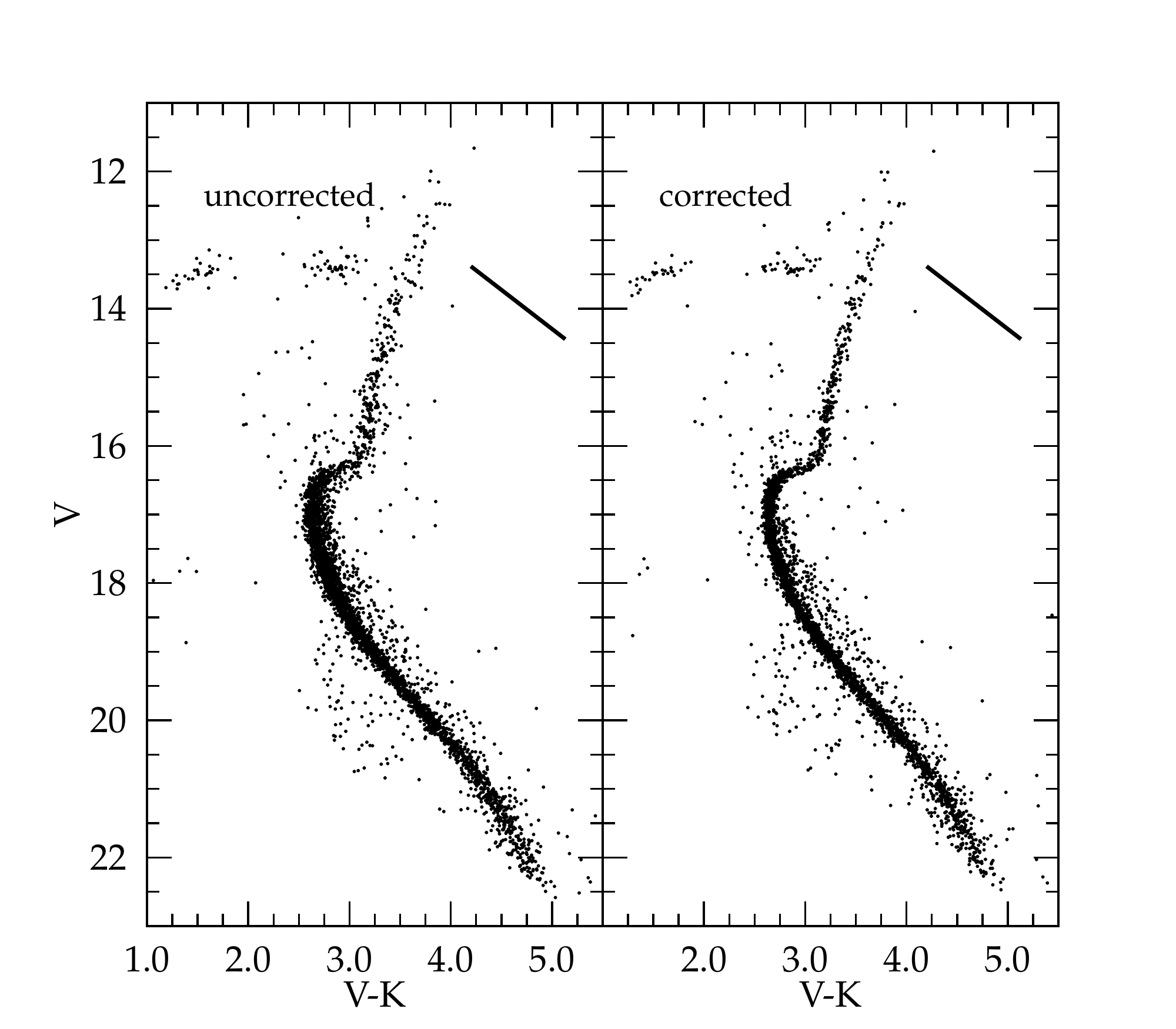}
\caption[Comparison between original and differential-reddening corrected CMD.] {Direct comparison of our HQ-sample of M4 before (left panel) and after (right panel) applying the derived differential reddening correction. Note especially, that the MSTO, the RGB, and the HB are significantly sharper in the corrected version of the CMD. This will help us to define the distance and age of M4 with a higher precision.
The black bar in each panel indicates the direction of the reddening vector and therefore defines the direction of the shift for stars with different individual extinctions. Its length reflects the shift caused by a total reddening of $E(B-V)=0.37$. }
\label{pre after1}
\end{center}
\end{figure}

\begin{figure}[t]
\begin{center}
\includegraphics[width=0.5\textwidth]{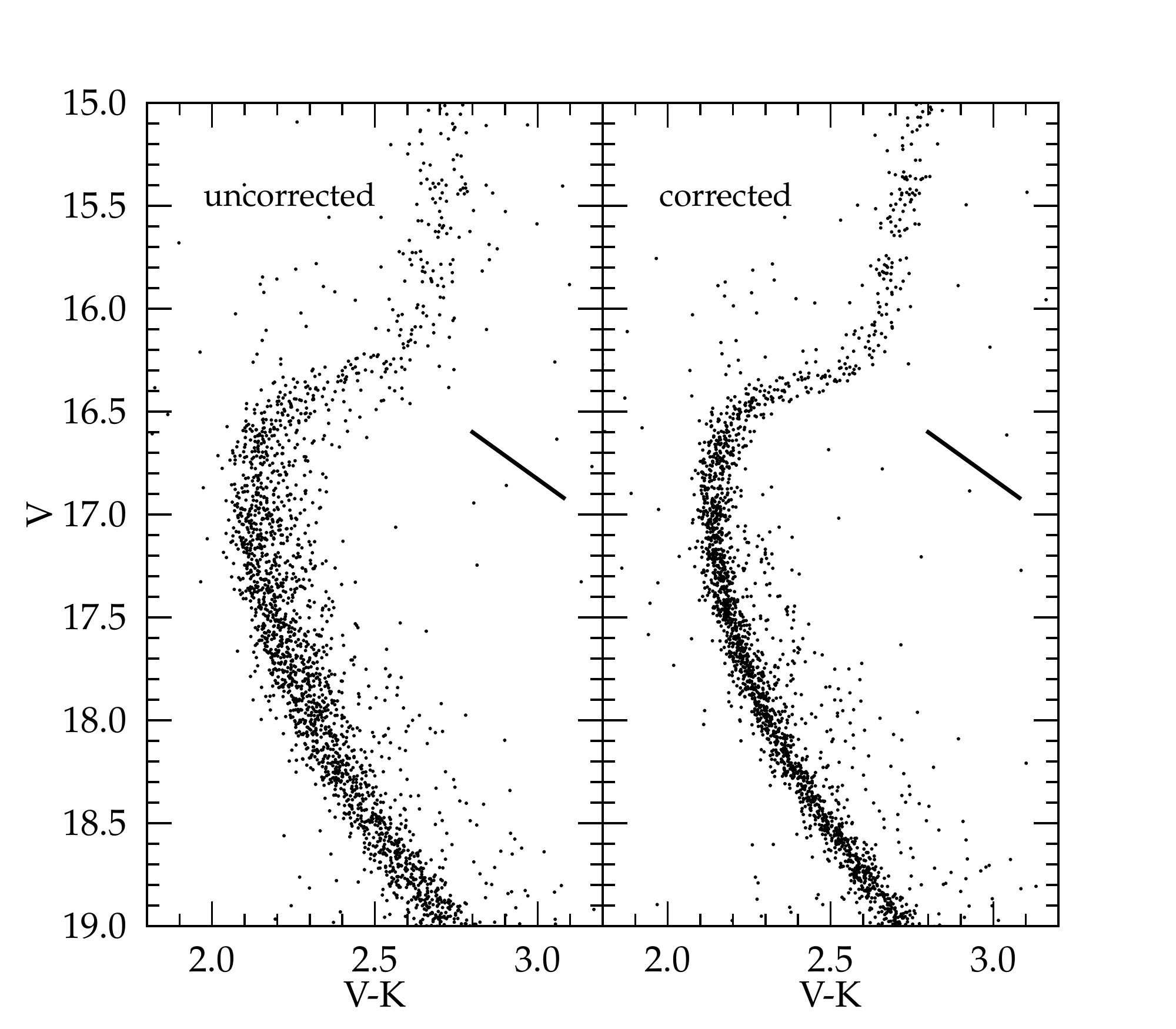}
\caption[Comparison between original and differential-reddening corrected MSTO section.]{Expanded view of the MSTO region before (left panel) and after (right panel) the differential reddening correction has been applied. We are able to decrease the observational scatter to about half of its original amount when we correct the photometry for spatial differential effects. As in the previous plot, the black bars indicate the direction of the reddening vector.}
\label{pre after2}
\end{center}
\end{figure}

Some of the spatially differential effects show strong evidence for systematic photometric errors within individual frames. In particular, the vertical feature at $x\approx0$ in Figure~\ref{reddening map} falls along the boundaries of the two most frequently observed regions. These systematic photometric errors are most likely the consequence of the PSF variations within SOFI that were mentioned previously. Alternatively, they could be the result of inadequate flat-fielding of the individual images or possibly crowding-dependent systematic errors in the 2MASS catalog itself (their angular resolution of $\sim 2^{\prime\prime}$ is appreciably poorer than that of our images).

Finally, we use the individual spatial differential reddening values to correct our photometry on a star-by-star basis. 
Figures~\ref{pre after1} and \ref{pre after2} show a comparison between our original CMDs and those corrected for differential reddening. With this correction, we are able to decrease the scatter in our data by about 50\%. The reduction in the scatter along the RGB and the HB is especially dramatic, which allows us to determine crucial parameters like the HB luminosity or the mean value of $E(B-V)$ with higher precision.

After the correction for differential reddening, we determine the fiducial sequence (including an empirical ZAHB sequence) for our later analysis of M4. For NGC 6723, corresponding information is assessed directly from its HQ-sample without an additional correction for spatial reddening variations. The fiducial sequences for both clusters can be found in Appendix \ref{appendix_b}.

\section{An Object-Specific Reddening Law}
\label{chapter_reddeninglaw}

A reddening law describes the functional dependence of interstellar extinction with observed wavelength. With increases in the total amount of extinction that an observed object suffers from, minor discrepancies in the applied corrections for reddening become more significant and can lead to incorrect implications for the distance or for basic atmospheric and chemical parameters derived from photometry.
Such discrepancies can arise either from inappropriate assumptions for the actual bandpass that is used (as defined by the instrument {\it and\/} the energy distribution of the observed flux) or from the applied reddening law as defined by the extinction properties of the dust in the line of sight to the object.
In this section we will focus on the extinction properties of our broadband filters for a given reddening law, while we will address the issue of the actual shape of the reddening law itself in section \ref{chapter_dusttype}. 

The extinction along the line of sight to a source is defined by how much the observed flux ($F_{\lambda}$) is suppressed relative to what it would have been if there were no dust ($F^0_{\lambda}$).
The fraction of the flux which reaches Earth, passes through the filter, and is measured by the detector is determined by the atmospheric transmission, the mirror reflectivity of the telescope, the transmission of the filter and other optical elements in the instrument, as well as the detector quantum efficiency.
The latter are \textit{source-independent factors} and, after accounting for the path length through the atmosphere, can usually be considered constant for a given observing site. Their product may be called \textit{system response function}, which we will designate $T_{\lambda}(\lambda)$. 

However, to calculate the extinction in a broadband filter ($A_{\Lambda}$), the energy distribution of the flux itself must also be taken into account. 
The spectral energy distribution ($S_{\lambda}(\lambda)$) reaching Earth from the observed object changes with \textit{source-dependent factors} such as temperature ($T_{\mathrm{eff}}$), surface gravity ($\log g$), metallicity ([Fe/H]) and the amount of extinction (e.g. $A_{V}$). Knowing the system response function and the individual stellar flux, the extinction for a given filter is given by

\begin{equation}
A_{\Lambda}=-2.5 \log \biggl({{F_{\lambda}} \over {F^0_{\lambda}}}\biggr),
\end{equation}
and therefore
\begin{equation}
A_{\Lambda}=-2.5 \log \biggl({{{\int\limits_{0}^{\infty}{D_{\lambda}(\lambda) T_{\lambda}(\lambda) S_{\lambda}(\lambda) 10^{-A_{\lambda}/2.5}\ d\lambda}}} \over {{\int\limits_{0}^{\infty}{D_{\lambda}(\lambda) T_{\lambda}(\lambda) S_{\lambda}(\lambda)\ d\lambda}}}}\biggr).
\label{extinction equation}
\end{equation}

\noindent $A_{\lambda}$ describes the strength of the extinction as a continuous function of wavelength and---in our case---the dust-type parameter $R_V$. $D_{\lambda}$ transforms the energy distribution of the spectrum from flux units to actual photon counts in the CCD and therefore is equal to $\lambda$ in our case.
Since $S_{\lambda}(\lambda)$ appears in both integrals, the physical constant involved in converting from energy to photon counts---$hc$ in $E={hc}/\lambda$---can be extracted from the integral and cancel out. In the next section we will specify the different components of Eq.~(\ref{extinction equation}) in detail. We use $\Lambda$ as a label representing some arbitrary broad photometric bandpass.

Practically, most reddening corrections are made by assuming a constant value for the extinction in broadband filters for every star in the sample calculated from an effective wavelength of the filter-detector system corresponding (in most cases) to a Vega-like flux distribution. However, this is only a first-order approximation to a more complex dependence on the aforementioned atmospheric and chemical parameters---notably the stellar surface temperature, as pointed out in former studies (e.g. \citealt{Bessell98}; \citealt{McCall04}; \citealt{Girardi08})---and is only valid if either the set of filters is sufficiently narrow, or the interstellar extinction is low, neither of which is the case for our data for M4.

Here, we want to calculate the precise total extinction in each of our optical and infrared filters, appropriate to the specific $T_{\mathrm{eff}}$, $\log g$ and [Fe/H] for each star in a system like M4. The correction factors so derived can be tailored to the relatively cool, metal deficient GC stars and we want to know how big the difference is compared to the standard extinction values provided in the literature (e.g. \citealt{McCall04}; \citealt{Schlegel98}; and \citealt{Cardelli89}), which are mostly derived from significantly hotter O-, B- or A-type stars.
Since the determination of the dust type depends upon the zero points for the transformation factors and consequently on the stellar properties, these results are necessary to provide appropriate zero points for the reddening law of M4 in order to minimize systematic errors in the determination of $R_V$ in the next section.
To calculate the actual extinction for each filter, we use Eq.~(\ref{extinction equation}) and vary the shape of the targeted SED ($S_{\lambda}$) and the reddening law ($A_{\lambda}$; see next section).

One important aspect of our study will be to compare the colour excess in different optical and NIR filters in order to derive a consistent reddening estimate, $E(B-V)$, for M4 from all filter combinations, or to explain the discrepancies. For this issue it is convenient to describe the reddening law in terms of the resultant transformation factors between the reddening in $(B-V)$ and any other filter combination $(\Lambda_1-\Lambda_2)$. Those factors can be calculated from the extinction values for each filter:

\begin {equation}
F_{\Lambda_1-\Lambda_2}={{A_{\Lambda_1}-A_{\Lambda_2}} \over {A_{B}-A_{V}}}. 
\label{E7}
\end {equation}

\subsection{Ingredients for the Reddening Law Calculations}
The NIR data in $J$ and $K_s$ are calibrated to stars of the 2MASS system. For this reason we use system response functions for the 2MASS filters provided in \citet{Cohen03}. These authors already combined all contributions from the optics, detector, location-specific atmosphere and filters so that we can use their so-called ``spectral response functions" directly as our $T_{\lambda}$. 

For the optical filters and the Landolt standard stars, such a sophisticated synthesis of detailed characteristics is not published; consequently we had to calculate the response function using Landolt's tabulated filter transmissions and photocathode sensitivity (\citealt{Landolt92}). We adopted the atmospheric transmission function from \citet[see p.~126]{Allen65} scaled to 0.7 atmospheres, approximately correct for an observatory 2~km above sea level. We further included two reflections from aluminum from \citet[see p.~108]{Allen65} as is appropriate for a Cassegrain instrument.
\label{reddeninglaw_ingredientsforthereddeninglaw}

One of the most crucial components of Eq.~(\ref{extinction equation}) is the dependence of the interstellar extinction on the wavelength. We are using the empirically derived analytical expression from \citet{Cardelli89} and specifically their equations (1), (2a), (2b), (3a), and (3b) to describe the extinction function. Since equation (1) has $R_{V}$ as a free parameter, we are able to derive extinctions for different types of dust.

With the use of different stellar spectra we are able to predict the reddening law for any combination of stellar temperature, surface gravity and metallicity, and investigate the differential effects for a stellar sample found---for example---within a globular or open cluster. The basic problem in the selection of stellar spectra is the difficulty of finding a database covering all wavelengths from near UV ($U$ filter at $\sim 3\,600$~\AA) to the $K_{s}$ filter at $\sim 21\,500$~\AA \ in the NIR.
In total we use three different spectral databases in our analysis, which should improve the reliability of our predictions: 

\begin{itemize}

\item The library of ATLAS9 model atmospheres (\citealt{Castelli03}) contains synthetic stellar fluxes for a fine grid in $T_{\mathrm{eff}}$ and $\log g$. Furthermore, it provides synthetic fluxes for stars of different metallicities from $\mbox{\hbox{\rm [Fe/H]}}=+0.2$ down to $\mbox{\hbox{\rm [Fe/H]}}=-4.0$.

\item The atlas from \citet{Pickles98} contains observed stellar fluxes, compiled by combining the data from several existing catalogues overlapping in wavelength coverage. The atlas provides 65 flux-calibrated spectra spanning a wavelength range of 1\,150 - 25\,000\,\AA \, and therefore encompasses all relevant optical and NIR filters.
Since we expect the reddening correction to depend on both dimensions (i.e., $T_{\mathrm{eff}}$ and $\log g$) of a CMD, we select a sample of spectra which follow a typical isochrone representation of a GC fiducial sequence as closely as possible. The vast majority of the spectra have solar metallicity ($\mbox{\hbox{\rm [Fe/H]}}=0.0$) so that we are not able to make predictions for metal deficient stars using this atlas. 

\item The MILES Library of Empirical Spectra (\citealt{Sanchez06}) provides observed spectra for stars with a wide range of metallicities and atmospheric parameters. We were unfortunately not able to select a sufficiently large sample of equal-abundance stars to occupy the whole evolutionary sequence of a GC. Furthermore, the wavelength coverage ranges from 3\,525~\AA \ to 7\,500~\AA \ and therefore fully covers only the $B$ and the $V$ filters. For this reason the MILES library is used only to empirically test the results for low metallicities as derived from synthetic spectra (see section \ref{reddeninglaw_effectsofmetallicity}). For this purpose we select a sample of spectra located close to the MSTO location in a CMD ($T_{\mathrm{eff}}\approx6\,000\,\mbox{K}$; $\log g\approx4.5$) with different metallicities typical of GCs.

\end{itemize}

\subsection{Results}
\label{reddeninglaw_results}
With the extinction law defined by Eq.~(\ref{extinction equation}) and the different components described in section \ref{reddeninglaw_ingredientsforthereddeninglaw}, we investigate the impact of different GC parameters on the reddening law which has to be applied to correct the data for interstellar reddening. In most cases we discuss the results using as an example the transformation factor $F_{V-K}$ between the observed reddening in $B-V$ and the derived colour excess in $V-K$. In this way, the impact on a CMD that combines optical and near infrared filters can be directly assessed, whereas a discussion on the basis of $A_{\Lambda}$ would keep the reader in the dark about the effects found when several extinctions are combined according to Eq.~(\ref{E7}).   
Throughout this section we assume a standard dust-type parameter of $R_{V}=3.1$. A higher value of $R_V$, however, would only cause a constant shift in $F_{\Lambda_1-\Lambda_2}$ but does not affect the relative changes caused by variations in stellar parameters.

Throughout this paper we will use the expression \textit{reddening-law zero point} for the value of $F_{\Lambda_1-\Lambda_2}$ or $A_{\Lambda}$ that is obtained for a star in the TO region of the CMD. It is therefore equated to the result of Eq.~(\ref{extinction equation}) for a star with atmospheric parameters $T_{\mathrm{eff}}=6\,000\,\mbox{K}$ and $\log g=4.5$. This definition has been adopted since, in most cases, this part of a GC evolutionary sequence is used to fit isochrone models to the data. In the case where the zero points are directly referred to M4, they further imply a metallicity of $\mbox{\hbox{\rm [Fe/H]}}=-1.0$. In contrast to that, we discuss the \textit{differential\/} effect in $F_{\Lambda_1-\Lambda_2}$ towards a given zero point due to variations in stellar parameters or---in the following section---$R_V$.

\subsubsection{Effects of Temperature}
\label{reddeninglaw_effectsoftemperature}
\begin{figure}[t]
\begin{center}
\includegraphics[width=0.5\textwidth]{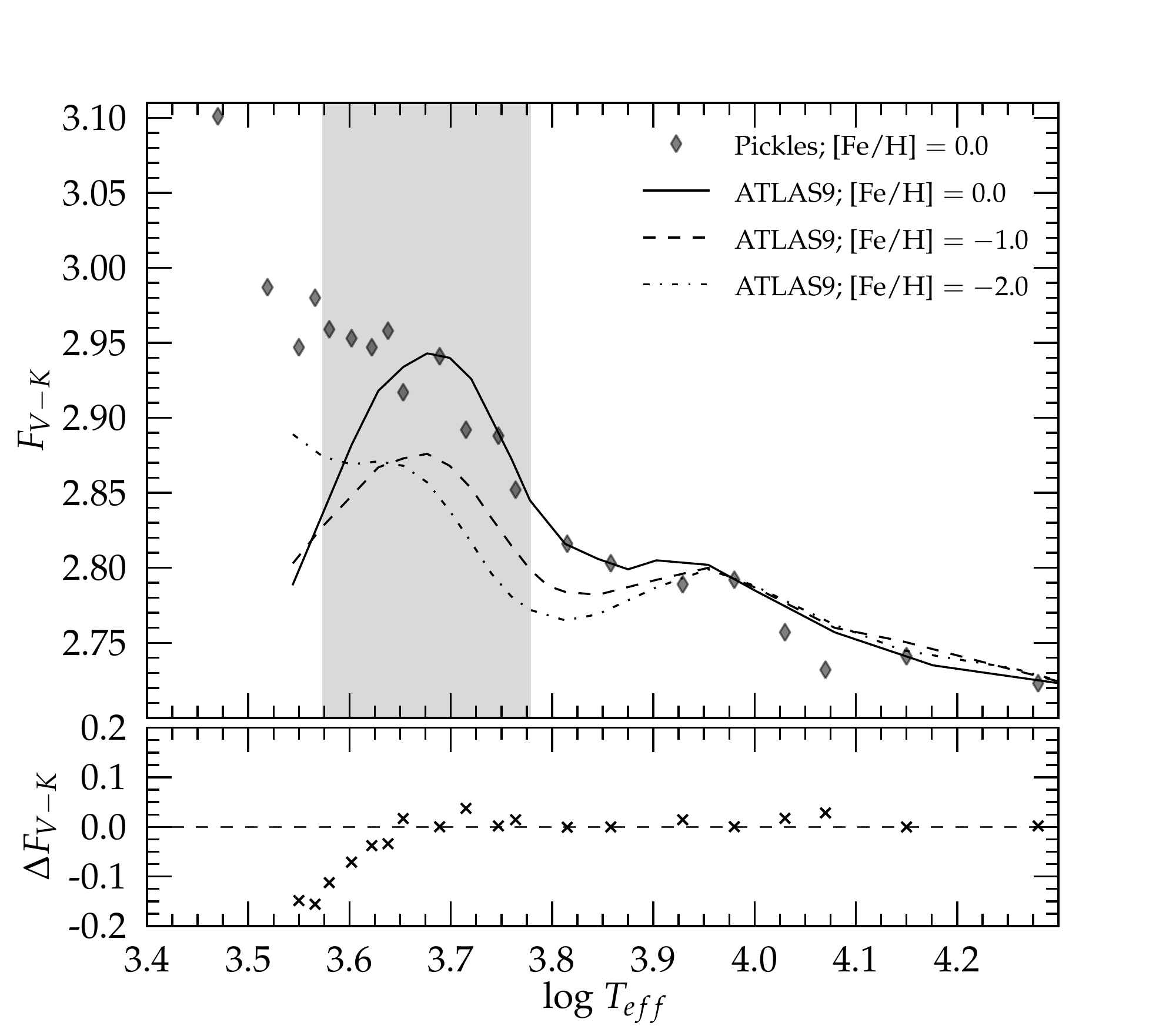}
\caption[Differential effects with temperature on the extinction correction.]{Dependence of the transformation factor $F_{V-K}$ on stellar temperature. In the upper panel, diamonds indicate spectra from the observed atlas from \citet{Pickles98} and black lines show the results obtained with synthetic ATLAS9 spectra from \citet{Castelli03}. The grey area indicate the range of typical GC temperatures with the lower boundary at temperatures for cool MS stars and the upper boundary at roughly TO temperatures for old stellar populations. The lower panel shows the differences between the models and real spectra at the solar metallicity. Good agreement is found for temperatures above $\sim5\,000\,\mbox{K}$.}
\label{T effect}
\end{center}
\end{figure}

Figure~\ref{T effect} shows the effect of varying the temperature on the transformation factor $F_{V-K}$. The results obtained with observed spectra are shown in direct comparison to those from synthetic spectra. For this comparison we choose a stellar sample from the Pickles atlas which approximately follows the MS of a GC evolutionary sequence, while for the ATLAS9 synthetic spectra we choose a constant $\log g=4.5$ and a wide range of temperature. 

Since we observe no significant differences in shape for the filter combinations relevant for this work (including $B$, $V$, $I$, $J$, $H$, or $K_s$), the most significant features of Figure~\ref{T effect} can be generalized to:

\newcounter{qcounter}
\begin{list}{\arabic{qcounter})~}{\usecounter{qcounter}}

\item The results obtained with the observed spectra show a relatively smooth increase for $F_{\Lambda_1-\Lambda_2}$ with decreasing temperature. Between the TO temperature and the coolest stars seen on the lower MS at $\log T_{\mathrm{eff}}\sim$~$3.58$, the change in $F_{\Lambda_1-\Lambda_2}$ is about 3\%.

\item The synthetic spectra match the results from the observed sample very well for temperatures above 5\,000\,K, whereas there is a strong deviation for lower temperatures. For $T_{\mathrm{eff}}\leq5\,000\,\mbox{K}$ the synthetic spectra display a strong decrease in the transformation factor after reaching a maximum value, whereas the observed spectra show a monotonic increase of $F_{\Lambda_1-\Lambda_2}$. 
This discrepancy seems to be stronger for more metal rich stars since the synthetic spectra with a metallicity of $\hbox{\rm [Fe/H]}=-2.0$ yield almost the same shape as the observed spectra. We further investigate the discrepancies between observed and synthetic spectra in the next section and discuss possible reasons for them.

\item Within temperatures ranging from 4\,500\,K to 6\,500\,K, a decrease in metallicity causes a significant decrease in the zero point of $F_{\Lambda_1-\Lambda_2}$ whereas the shape or the differential effect is almost the same. For $T_{\mathrm{eff}}\geq9\,000\,\mbox{K}$ almost identical values of $F_{\Lambda_1-\Lambda_2}$ are found for all metallicities: metallic features have no perceptible effect on stellar spectral energy distributions at these temperatures. 

\item Principally, our zero points for the transformation factors which include $J$, $H$, or $K_s$ filters seem to be higher then the ones commonly used in the literature. For example, $F_{V-K}=2.78-2.85$, depending on the metallicity, whereas the canonical value using a Vega-like star is about 2.71 (\citealt{McCall04}). 

\end{list}

\subsubsection{Discrepancies between observed and synthetic spectra}

\begin{figure}[t]
\begin{center}
\includegraphics[width=0.5\textwidth]{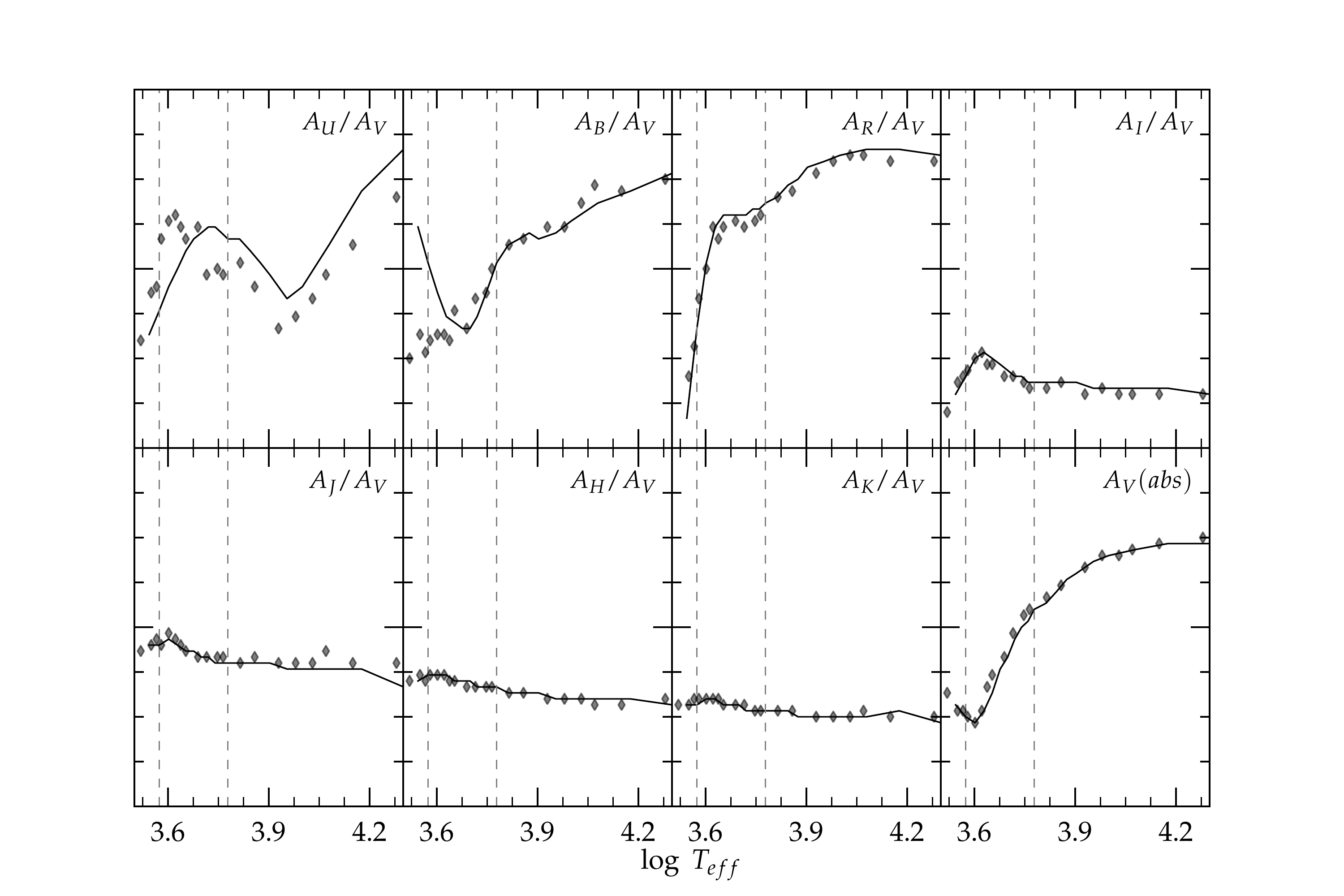}
\caption[Differential effects with temperature on the individual filter passbands.]{Relative extinctions for all filters as derived from the observed spectra from \citet{Pickles98} (diamonds) and synthetic ATLAS9 spectra (solid line) from \citet{Castelli03}: Although the actual numbers are different, the $y$-axes show the same scale for each subplot ($A_{\Lambda}/A_{V}=0.0075$ between tickmarks) so that relative changes with temperature and differences between synthetic and observed spectra can be compared directly. There is good agreement for all photometric filters, including the absolute value for the $V$-band. Only the extinction in the $B$-filter shows significant differences between the different types of spectra for temperatures below 5\,000\,K. The range of typical GC temperatures lie between the dashed vertical lines.}
\label{single extinctions}
\end{center}
\end{figure}

According to Eq.~(\ref{E7}), $F_{\Lambda_1-\Lambda_2}$ includes the extinction for at least three different filters. To find the source of the discrepancy between the models and the real spectra in Figure~\ref{T effect}, we looked into each component separately and compared the extinction in all filters for the observed and the synthetic spectra to determine whether there are general differences or whether they are related to specific filters. The results are shown in Figure~\ref{single extinctions}. We find that there is very good agreement for each filter except $U$ and, more dramatically, $B$ which explains very well the deviation between the Pickles and ATLAS9 spectra in Figure~\ref{T effect} because a higher value of $A_{B}$ results in a lower value of $F_{V-K}$. Since $A_{B}$ contributes to every transformation factor, it can explain the observed deviations for all filter combinations.

\begin{figure}[t]
\begin{center}
\includegraphics[width=0.5\textwidth]{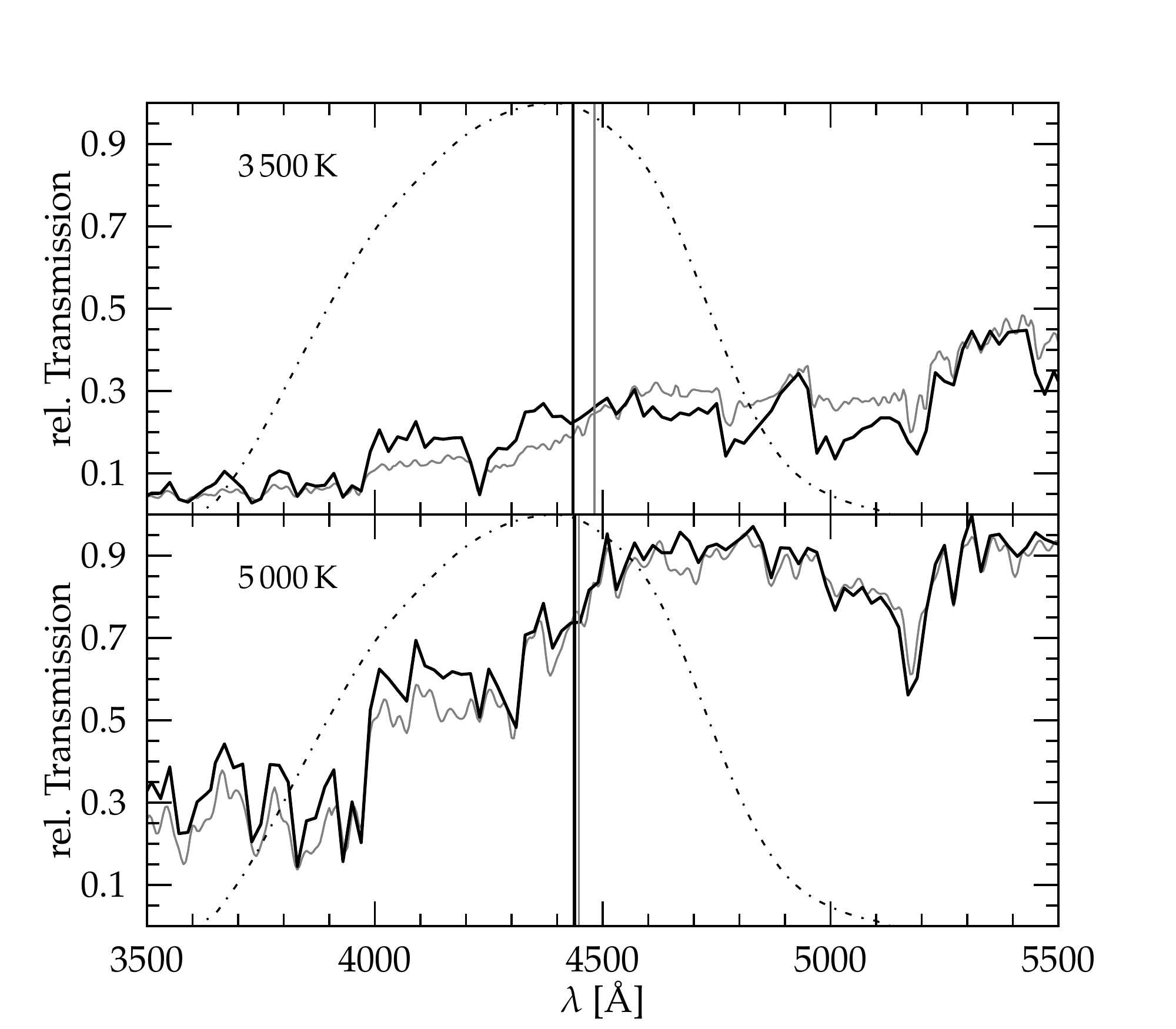}
\caption[Comparison between synthetic and observed flux.]{Detailed plot of the different spectral energy distributions at $T_{\mathrm{eff}}=3\,500\,\mbox{K}$ (top) and $T_{\mathrm{eff}}=5\,000\,\mbox{K}$ (bottom) to visualize the cause of the extinction differences in the $B$-band. Vertical lines indicate the resultant effective wavelength for $B$ when the different spectra are used (thick black: synthetic ATLAS9; grey: observed spectrum). The significantly shorter effective wavelength for ATLAS9 spectra in the top panel results in the higher relative extinction in $B$ compared to observed spectra. For $T_{\mathrm{eff}}=5\,000\,\mbox{K}$. The differences in the spectral flux are much smaller and the effective wavelength is basically the same for synthetic and real spectra. Both spectra are normalized to the same wavelength so that intensity differences can be compared directly.}
\label{b filter view 1}
\end{center}
\end{figure}

To explain the difference in the spectral energy distributions, we overplotted a real spectrum with the synthetic spectrum in the wavelength range of the $B$ filter for $T_{\mathrm{eff}}=3\,500\,\mbox{K}$, where we expect a strong deviation, and for $T_{\mathrm{eff}}=5\,000\,\mbox{K}$ where we expect little or no deviation (see Figure~\ref{b filter view 1}).
In the cooler star, the synthetic atmosphere models (black curve) consistently {\it over\/}estimate the true stellar flux (grey curve) by many percent at the shorter wavelengths within the bandpass, while they consistently {\it under\/}estimate the flux in the longward portion of the bandpass. This causes a shift in the photon-weighted mean wavelength of the light that passes through the filter (vertical lines), which results in a mis-estimate of the net extinction integrated over the $B$ bandpass. 

For low temperatures, molecular lines and bands become more prominent in the optical part of the spectrum, including the wavelength range of the $B$ filter. Specifically, the wavelength range encompassing the observed discrepancies includes prominent CN and CH molecular bands which are widely used as spectral indices to investigate variations in the CNO-cycle elements in GCs (\citealt{Norris79}). The indices S3839, S4142 (for CN) and CH~4300 (for CH) as defined in \citet{Harbeck03} are located exactly in the region where we observe the strongest deviations (see e.g., \citealt[Fig.~3]{Pancino10}). 
Apparently, the synthetic atmosphere models do not perfectly reproduce these complex features (e.g., \citealt{Bessell98}). This fact, together with the increasing deviation towards lower temperatures and higher [Fe/H] supports the assumption that the results obtained with the synthetic ATLAS9 spectra are not entirely trustworthy for the cool temperature regime due to an inadequate treatment of molecular bands.  

\newpage
\subsubsection{Effects of Metallicity}
\label{reddeninglaw_effectsofmetallicity}
Since the different spectral samples agree very well for temperatures above 5\,000\,K, we infer that the synthetic models predict the change of $F_{\Lambda_1-\Lambda_2}$ zero points with metallicity correctly at MSTO temperatures. To confirm this inference, we use the spectral atlas from \citet{Sanchez06} to compare the results of the ATLAS9 models with the observed spectra for a wide range of metallicity between $\hbox{\rm [Fe/H]}=+0.2$ and $\hbox{\rm [Fe/H]}=-2.5$ at a constant temperature of $T_{\mathrm{eff}}\approx6\,000\,\mbox{K}$.

Unfortunately, we are not able to generate extinctions for each filter with the observed metal poor spectra, due to their limited wavelength coverage. For this reason we derive transformation factors where the observed metal poor spectra are used only for the $B$ and $V$ filters, whereas we derive the extinction for the remaining bands with synthetic metal poor spectra. Then, we compare this hybrid model to a purely synthetic one where the extinctions for all filters are derived from synthetic spectra. Note that for $F_{V-K}$, for example, only the extinction in the $K_s$ filter is calculated with synthetic fluxes.
Since we expect changes in the metallicity will mainly impact the optical part of the spectrum (and therefore $U$, $B$, and $V$), we are confident that the results obtained with the hybrid model reflect the true situation very well. 

\begin{figure}[b]
\begin{center}
\includegraphics[width=0.5\textwidth]{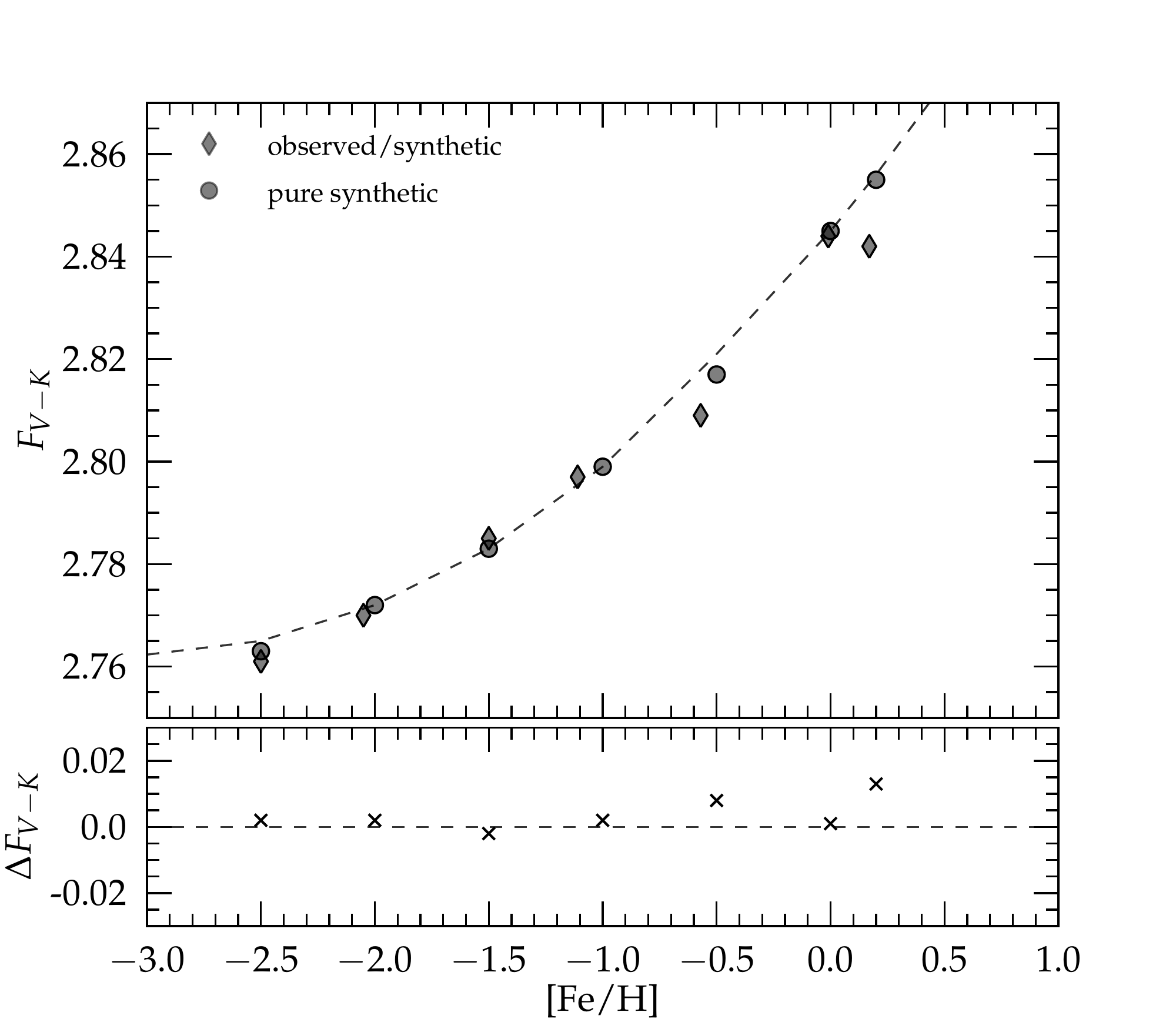}
\caption[Differential effects with metallicity on the extinction correction factor.]{Top: Zero points for $F_{V-K}$ as a function of \hbox{\rm [Fe/H]} for the hybrid combination (diamonds) and a pure synthetic sample (circles) are compared directly. The dashed line indicates the functional dependence of $F_{V-K}$ with \hbox{\rm [Fe/H]} as derived from pure synthetic spectra with $T_{\mathrm{eff}}=~6\,000\,\mbox{K}$ and $\log g=4.5$ which represents the atmospheric parameters at the MSTO. Note that these parameters are not used for all ATLAS9 spectra in the direct comparison sample (see Table~\ref{syn-obs-spectra}) and therefore some circles do not fall exactly on the dashed line.
Bottom: Deviation in $F_{V-K}$ between pure synthetic and hybrid metal poor spectra. $\Delta F_{V-K}~=~F_{V-K}(syn) - F_{V-K}(hyb)$.}
\label{metallicity result}
\end{center}
\end{figure}

Table~\ref{syn-obs-spectra} summarizes the parameters of S\'anchez-Bl\'azquez and the ATLAS9 spectra which are used for the direct comparison. Figure~\ref{metallicity result} shows the derived values for $F_{V-K}$ using the synthetic and the hybrid model, together with the differences between them. 
The consistency between synthetic metal poor spectra and the observed ones is extremely good. Only the results from the most metal rich synthetic spectra for $\hbox{\rm [Fe/H]}\geq 0.0$ deviate from the values obtained with observed fluxes. For the rest of the direct comparisons the difference in $F_{V-K}$ is considerably smaller than the photometric uncertainty without any systematic deviation. 

Reviewing Figure~\ref{T effect}, we want to point out that the solar-metallicity synthetic spectra reproduce very well the shape of all $F_{\Lambda_1-\Lambda_2}$ derived from observed spectra above 5\,000\,K. By comparing synthetic spectra with different metallicities, we find the shape to be very similar for all metallicities in the temperature range $5\,000\,\mbox{K} \leq T_{\mathrm{eff}} \leq 9\,000\,\mbox{K}$. Therefore, a change in metallicity in this temperature range mainly produces a shift in the zero-point value of the transformation factor. As a result, we can calculate these zero points from metal poor synthetic spectra in a temperature range where the results are confirmed by metal poor observed spectra and then apply them to different temperatures. However, it is important to point out that the effect of metallicity on the reddening law is not constant over larger temperature scales and the use of a constant zero-point offset is an approximation that seems to be justified only for the limited range in GC temperatures.   

\begin{table}[b]
\centering
\caption{Parameters for observed and synthetic spectra in direct comparison.}
\begin{tabular}{rcccrcc}
\hline
\hline
                		      &Sanchez-B.			& 	&	&			&ATLAS9&					\\
$\hbox{\rm [Fe/H]}$		&	$\log T_{\mathrm{eff}} $	& $\log g$	&&  $ \hbox{\rm [Fe/H]}$		&	$\log T_{\mathrm{eff}}$	&$\log g$\\\cline{1-3}\cline{5-7}
  +0.17	   	&	3.777 		& 4.02	&&	+0.20		&	3.778		&	4.00	      	\\
   -0.01    		& 	3.772		& 4.10	&&	+0.00		&	3.778		&	4.00		\\
   -0.57		&	3.779		& 3.99	&&	-0.50		&	3.778		&	4.00		\\
-1.11		&	3.767		& 4.45	&&	-1.00		&	3.778		&	4.50		\\	
-1.50		&	3.765		& 4.30	&&	-1.50		&	3.778		&	4.50		\\
-2.05		&	3.772		& 4.35	&&	-2.00		&	3.778		&	4.50		\\
-2.50		&	3.786		& 2.60	&&	-2.50		&	3.778		&	2.50		\\\hline
\end{tabular}
\label{syn-obs-spectra}
\end{table}

\newpage
\subsubsection{Effects of Gravity}
\begin{figure}[t]
\begin{center}
\includegraphics[width=0.5\textwidth]{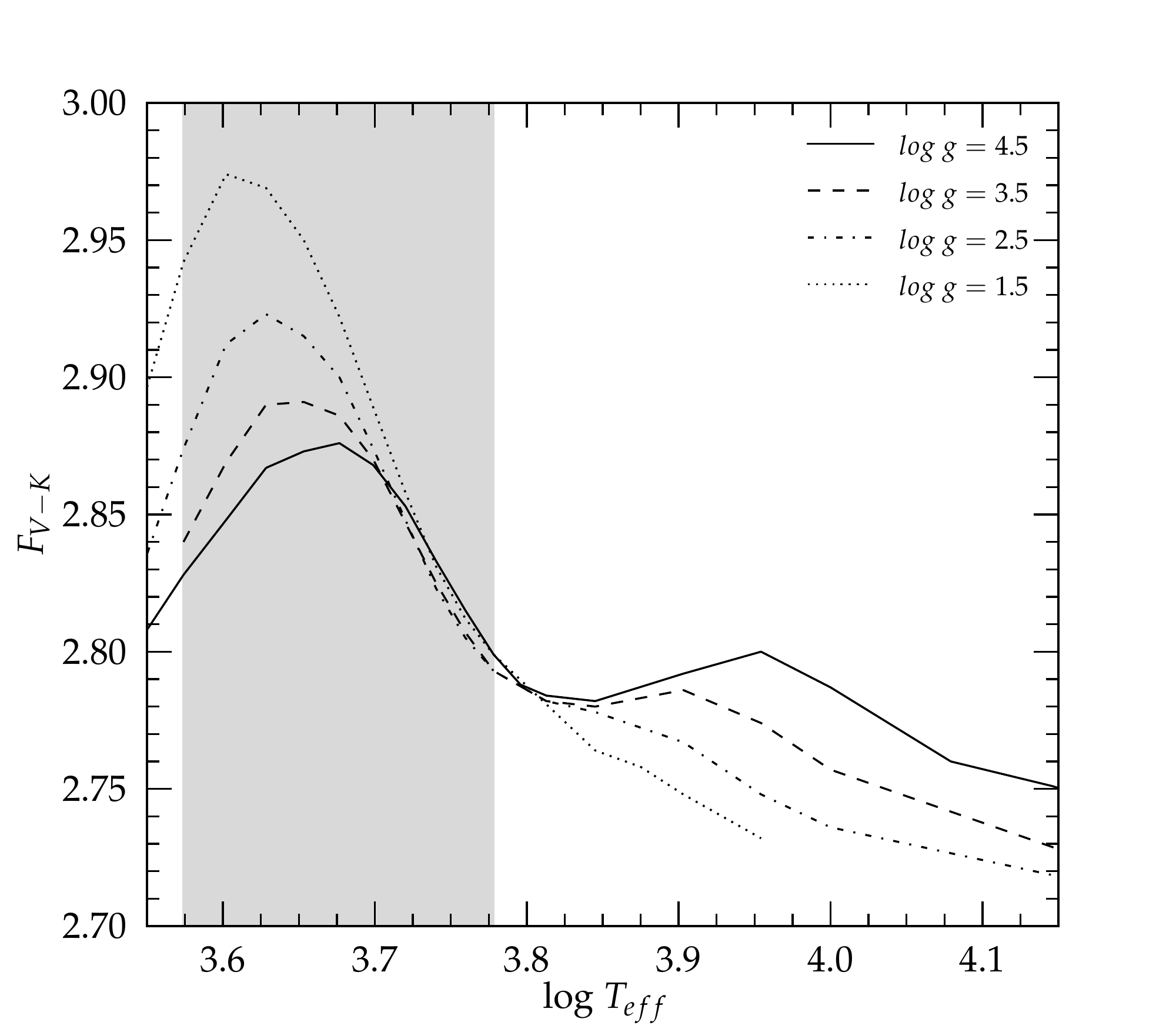}
\caption[Differential effects with surface gravity on the extinction correction factor.]{Effects of surface gravity on the transformation factor $F_{V-K}$ for a wide range of temperatures. For this calculation we used synthetic spectra with $\mbox{[Fe/H]}=-1.0$ and [$\alpha$/Fe]$=+0.4$ to simulate the chemical composition of the stars in M4. As in previous plots, the grey area indicates the range of typical GC temperatures. In general, a lower surface gravity results in higher values for $F_{V-K}$, where this effect seem to increase with lower temperatures and is not observable for temperatures near the MSTO. The effect seems to be inverted for temperatures hotter than $T_{\mathrm{eff}}\approx6\,300\,\mbox{K}$.}
\label{gravity effect}
\end{center}
\end{figure}

The absolute magnitude $M_V$ of a star in a simple stellar population such as a GC system is strongly correlated with its surface gravity and therefore has a possible impact on the stellar spectrum and the reddening law which has to be applied. Figure~\ref{gravity effect} shows the effect of different $\log g$ values on the transformation factors derived with synthetic spectra from ATLAS9. Ignoring the temperature range cooler than the peak in $F_{V-K}$, we observe an increase for $F_{\Lambda_1-\Lambda_2}$ with decreasing surface gravity. The effect becomes more significant towards cooler temperatures, where it is of the same order as the temperature effect itself. For the lowest GC temperatures, the change in $F_{\Lambda_1-\Lambda_2}$ between a dwarf and a giant is about $3\%$ for all transformation factors. This is especially important for stars on the RGB since the decrease in temperature and surface gravity combine and enhance the differential effect of reddening along the RGB sequence. 
However, for temperatures between 5\,600\,K and 6\,300\,K the differential effect with gravity practically disappears and eventually reverses for even higher temperatures.

\subsubsection{Effects of Extinction}
\begin{figure}[t]
\begin{center}
\includegraphics[width=0.5\textwidth]{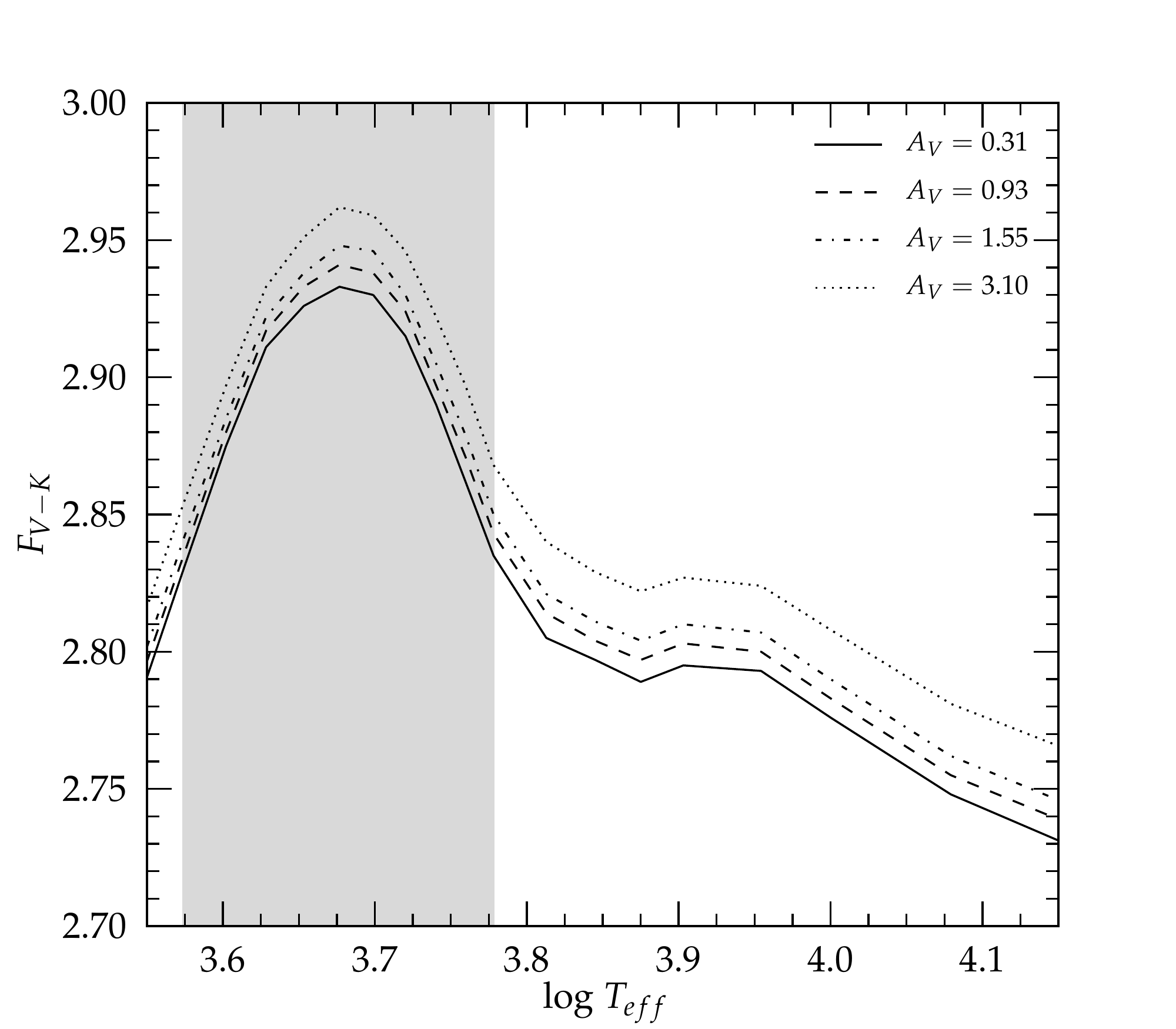}
\caption[Differential effects with extinction strength on the extinction correction factor.] {Effects of total extinction for four different total extinctions corresponding to $E(B-V)=0.1$, 0.3, 0.5 and 1.0 assuming $R_V=3.1$. The effect is small compared to atmospheric influences that have been discussed in the previous sections and has about the same impact for all temperatures. The grey area indicates the range of typical GC temperatures.}
\label{extinction effect}
\end{center}
\end{figure}

The total strength of interstellar extinction itself is a factor which impacts on the effective wavelength of the filters. This is because interstellar extinction is a function of wavelength and therefore changes the distribution of the incoming flux by damping the blue part of each bandpass more strongly than the red part. 
Figure~\ref{extinction effect} shows the temperature-dependent value of $F_{V-K}$ for different absolute extinctions $A_{V}$. Compared to atmospheric parameters, the influence of $A_{V}$ on the transformation factor is very small and although we use $E(B-V)=0.36$ in our calculations to tailor the results for M4, they are basically valid for at least all $E(B-V)\leq0.5$.
 
\subsubsection{Star-by-Star Variations within a Stellar Population}
Despite the definition of appropriate zero-point values for relative extinctions $A_{\Lambda}/A_{V}$ and transformation factors $F_{\Lambda_1-\Lambda_2}$ appropriate to the atmospheric and chemical conditions of GC stars, the question comes up whether changes of those parameters \textit{within} a typical GC population cause any significant change in the appearance of the cluster fiducial sequence compared to a constant correction with respect to the photometric uncertainty. Such changes could affect the determination of [Fe/H], distance, age and other parameters when determined from photometric studies.

For that reason we compared the extinction for a faint MS star ($T_{\mathrm{eff}}\approx3\,680\,\mbox{K}$, $M_V\approx9.05$) to a TO star ($T_{\mathrm{eff}}\approx5\,810\,\mbox{K}$, $M_V\approx4.25$) and a star on the upper RGB ($T_{\mathrm{eff}}\approx3\,820\,\mbox{K}$, $M_V=-0.66$) to simulate the highest possible differential effects in both temperature and surface gravity for a reddening of $E(B-V)=0.36$. 
Using SEDs with appropriate parameters, we found differential effects of $\Delta E(V-K)\approx0.025$~mag between the MS star and the TO star and $\Delta E(V-K)\approx0.05$~mag between the RGB star and the TO star. This effect is small and, for both examples, of the same order as or even below the photometric uncertainty. Therefore the \textit{shape} of the evolutionary sequence of clusters with significant reddening such as M4 are not affected significantly by differential effects due to variations in temperature and surface gravity. That means that parameters like metallicity and age---when derived by isochrone fitting---are not significantly biased by reddening zero point variations within the population. 
Specifically, the luminosity of the HB, which is often used to derive the distance modulus of GCs, is not biased when the zero-points are normalized to TO temperatures; at those temperatures, the change in surface gravity does not affect the reddening zero points. In general, however, the error caused by the use of a constant reddening correction is of systematic nature and a star-by-star correction should be considered when the highest possible photometric precision is desired or if the reddening in the line of sight is larger than for M4. 
In Appendix \ref{appendix_b}, we therefore provide two fiducial sequences of M4: one for the (apparent) observed sequence and one where the aforementioned differential effects are corrected on a star-by-star basis with respect to MSTO conditions.

\subsubsection{Extinction Zero Points}
\label{reddeninglaw_extinctionzeropoints}

From our calculations, we provide zero points for relative extinction values ($A_{\Lambda}/A_{V}$) for a wide range of different temperatures, metallicities, and values of $R_V$. From them, appropriate colour transformation factors can be derived. 
The metallicity and $R_V$ tables have been created with synthetic ATLAS9 spectra and observed fluxes are used to calculate values for the temperature table due to the uncertain treatment of molecular absorption in the models. For all tables, we assume $A_{V}=1.12$ (suitable for M4), but the values are valid for practically all $ A_{V} \leq 2$. These transformation factors should be applied to correct the colour excesses in filter combinations of optical Johnson-Cousins and/or NIR 2MASS filters. 
Since the tabulated differential effects due to temperature and [Fe/H] are basically independent of $R_V$, the tabulated offset due to the dust type can be used for any combination of the atmospheric and chemical parameters. The same can be assumed for the $T_{\mathrm{eff}}$ and [Fe/H] effects in the range between 5\,000\,K and 9\,000\,K.

The extinction tables are listed in Appendix \ref{appendix_c}, where they are compared to prominent literature values. Note that for $J$, $H$, and $K_s$, only \citet{McCall04} uses the actual 2MASS bandpasses, whereas all other references use slightly different filters.
 
\subsection{Comparison to existing Work}
\label{reddeninglaw_comparisontoexistingwork}
Some studies have considered the effects of temperature, surface gravity or other parameters on the reddening law. However, most of them use only synthetic spectra. 
The recent work from \citet{Girardi08} (hereafter G08) uses ATLAS9 spectra (the same generation than we adopt) to produce a differential reddening law with a grid of $\log g$ and $T_{\mathrm{eff}}$ specifically applicable for the WFPC2 and ACS photometric systems from the Hubble Space Telescope. They found a differential effect of $\Delta(colour)\sim 0.05\ \mbox{mag}$ for a total hypothetical extinction of $A_V \approx 3.0\ \mbox{mag}$, which corresponds to a reddening of $E(B-V) \approx 1.0$ using a standard $R_V$ value of $3.1$. We find a similar sized effect for less than half the reddening using a $V-K$ magnitude plane, which can be explained with the larger wavelength range that is spanned by this filter combination. All values from G08 are derived with solar metallicity spectra only; they claim that the variation with metallicity is negligible. That is not what we find in our study: even if the metallicity effect for the absolute extinction is very small in a single filter, it can increase significantly when combining four filters to derive the transformation factors between different colours. Using the same thresholds for [Fe/H] as G08, we find a difference of $\delta F_{V-K} \approx 0.08 \ \mbox{mag}$ between $\mbox{\hbox{\rm [Fe/H]}}=0.0$ and $\mbox{\hbox{\rm [Fe/H]}}=-2.5$, a change of $2.8$\% which is about ten times larger than G08 states. Overall, our work reveals a roughly equally sized effect for changes in [Fe/H], $T_{\mathrm{eff}}$, and $\log g$ ($\sim 3$\% each) encompassed by GC stars, whereas G08 claim that the effects of temperature variations dominate over those of surface gravity while those arising from metallicity variations are negligible for the HST filters.

Another interesting issue to discuss is the quality of ATLAS9 spectra, especially for low-temperature stars where molecular lines become important. Our work finds very good agreement between observed and synthetic spectra for all bands except $B$ for $T_{\mathrm{eff}} \leq 5\,000\,\mbox{K}$. Interestingly, \citet{Bessell98} use an older version of ATLAS9 models (from 1993) and found a good fit only for temperatures above $4\,500\,\mbox{K}$ due to their incomplete molecular opacity at lower temperatures. Although we are using a more recent version (2003) with an improved treatment of molecules, we seem to find a similar effect setting in at almost the same temperature. Carefully examining Fig.~2 in G08 indicates that their study suffers from the same problem: the relative extinction $A_{\Lambda} / A_{V}$ ($A_{S_{\lambda}} / A_{V}$ in G08) for the F435W filter---which is closest to Johnson $B$---shows a dip at cooler temperatures similar to our results with ATLAS9, whereas the use of observed spectra reveals a monotonic decrease in relative extinction towards lower temperatures.      

The work from \citet{McCall04} is primarily focused on the extinction correction of integrated SEDs for extragalactic sources, suitable---for example---for the determination of accurate distances from Cepheids. However, the author discusses the effect of temperature for resolved stellar spectra as well, using the \citet{Pickles98} dataset and found differences up to 23\% between unevolved O5 and M6 stars at the solar metallicity and a nearly equal effect between very cool evolved and unevolved stars, which is consistent with what we find when we consider the full range of temperature. Note that the largest effect is caused only by the coolest stars with $T_{\mathrm{eff}}\leq3\,500\,\mbox{K}$ and therefore differences are much smaller if limited to the range of GC temperatures. Unfortunately, McCall only provides zero-point values normalized to a Vega-like flux together with $R_V=3.07$ for a ``normal'' reddening law, which makes it difficult to apply his results to globular cluster conditions and compare them to our results in more detail. 

\section{The Dust Type of M4}
\label{chapter_dusttype}

The type of dust that causes the interstellar reddening can show significant variations for different lines of sight and is responsible for significant changes in the properties of absorption and scattering. Therefore the actual reddening law is mainly determined by the specific properties of the dust. This makes it, in principal, necessary to determine an individual reddening law for each object that suffers significant interstellar extinction, before correcting for those effects appropriately without inducing systematic errors.
To good approximation, the dependence on the dust type can be parameterized by the ratio between total-to-selective extinction, $R_V\equiv A_{V}/E(B-V)$, which seems to be strongly correlated with the grain properties of the dust (\citealt{Cardelli88}, \citealt{Mathis90}) and varies between $R_{V}\approx2.6$ and a value up to $\sim6.0$ for dense, cold molecular clouds or star forming regions. Since it defines the shape of the reddening law and the ratio between extinction and reddening, a precise knowledge of $R_V$ (or some equivalent parameter) is crucial for the absolute distance determination of stellar systems.
Usually a value of $\sim3.1$ is used to describe the foreground reddening for GCs in the Milky Way as it is associated with the average extinction properties of the ISM. However, there is strong evidence in the literature that M4 follows a different, anomalous reddening law due to the fact that the cluster lies behind the Sco-Oph dark cloud complex which adds an additional uncertainty to the distance and possibly to the age of the cluster as well as to the assumed mean reddening expressed in $E(B-V)$.

In this section, we want to determine the properties of the dust in the line of sight to M4, characterized by the appropriate value of $R_V$ in the reddening law given in \citet{Cardelli89}. From this, the actual ratio between $A_{V}$ and $E(B-V)$ for our photometry can be determined in order to derive the distance to the system.

\subsection{Method}
\label{dusttype_method}
In general, we want to find out which value of the dust type parameter $R_V$ produces a consistent match for isochrones with the observed CMDs of M4 in all filter combinations once the empirical offset is given for one colour. Here, we make use of the possibility of treating the dust type as a free input parameter to calculate transformation factors $F_{\Lambda_1-\Lambda_2}$ between different colours as a function of $R_V$. The dust type parameter enters equation (\ref{extinction equation}) twice: First, it defines the slope of the reddening law according to the Cardelli et~al.~equations. Second, it defines the actual extinction $A_{V}$ for a given reddening $E(B-V)$.

As far as we are aware, the only similar attempt to determine the ratio of total to selective extinction in the line of sight to star clusters can be found in \citet{An07}. Here, $R_V$ is determined by a $\chi^2$ minimization of the \textit{shape} of the reddening law defined by the observed colour excesses in optical and NIR broadband filters. 
In contrast, we choose one reference colour ($B-V$) and find the best value of $R_V$ by matching all other filter combinations \textit{individually}.
Compared to \citet{An07}, our approach is more sensitive to variations in the assumed value of [Fe/H], mainly because metallicity changes the theoretical colours for the isochrones and therefore the adopted value of $E(B-V)$. On the other hand, we are able to minimize filter-specific systematic errors for our bandpasses, while a $\chi^2$ minimization of the overall shape does not take those distorting effects into account and therefore can lead to systematically wrong answers if in fact the theoretical colour for only one bandpass is inaccurate. 
Since we have two powerful tools to constrain the metallicity of M4 (spectroscopic data and CMD isochrone fitting) but almost no access to the accuracy of synthetic filter colours, we decided not to use a minimization algorithm to determine $R_V$. 
Furthermore, An et~al. indeed account for temperature effects in the broadband filter extinction properties by using the colour-correction term provided by \citet{Bessell98}, but they do not consider any adjustment for gravity or metallicity as we do in our study.

\subsubsection{Empirical Color Offsets}
To determine the empirical colour offset between the observed CMD and the theoretical isochrone for each combination of filters, we use an age-independent fiducial point on the MS, defined to be 5.5 mag below the red ZAHB $V$-band magnitude. For M4, this occurs at $V=18.96$. Our fiducial point is insensitive to age, since it lies well below the MSTO and it is further defined with respect to the ZAHB luminosity, a standard candle which is known to be nearly independent of age (e.g., \citealt{Stetson96}). This is important since ages are not precisely known for GCs and could consequently cause a systematic error in the determination of $R_V$. We further choose the fiducial point intentionally below the MSTO to avoid known uncertainties within theoretical isochrones for evolutionary stages later than the TO due to the onset of---yet poorly understood---deep mixing processes (see, e.g., \citealt{VandenBerg05}).

We find the colour of the fiducial point in the observed CMD by calculating fiducial sequences along the MS of M4 in each desired filter combination using data that have been corrected for spatial differential reddening.
Fiducial points are obtained by binning the data in magnitude intervals of 0.3 along the MS and calculating the median value for each bin. The fiducial sequence is then defined by cubic spline interpolation between the fiducial points.    
This approach has the advantage of being consistent for all the colours considered: since both the reference luminosity and the colour of the fiducial point are derived the same way for all filter combinations, any systematic uncertainty cancels out to first order. In other words, even if the derived fiducial sequence might not represent the actual median reddening value for a given colour (e.g., due to a bias caused by binary stars), it still does represent \textit{the same} reddening value for all colours. Furthermore, using the median value to define the fiducial points, we minimize the influence of outliers such as field stars, blended stars and specifically binary stars that do not fall on the actual MS.  

Finally, each colour offset is given by the colour difference between the empirical fiducial point and its theoretical equivalent on the appropriate isochrone. 

From the individual observed colour excesses we derive the necessary transformation factors $F_{\Lambda_1-\Lambda_2}$ for each filter combination and determine the corresponding $R_V$ values from our reddening law calculations. The zero points of the transformation factors are calculated for $\mbox{\hbox{\rm [Fe/H]}}=-1.0$ (see section \ref{dusttype_isochrones} for a discussion about the metallicity of M4), $E(B-V)=0.37$ and for the atmospheric parameters $T_{\mathrm{eff}}=5\,250\,\mbox{K}$ and $\log g=4.5$ based on the location in the CMD of the fiducial point that is used to determine the offset.

\subsubsection{Error Estimation}
\label{dusttype_error_estimation}
In addition to the best fitting values of $R_V$, we estimate for each colour the observational standard error of the derived offset to the isochrone $(\sigma(\Delta colour))$ and the uncertainty in $R_V$ ($\sigma(R_V)$) which defines the overall precision of our method. Here, we have to take into account the photometric uncertainty of both the observed ZAHB luminosity ($\sigma (V_{HB})$) and the colour of the fiducial point at that luminosity ($\sigma(colour)$):

\begin{equation}
{\sigma^2(\Delta colour)}={\sigma^2(V_{HB})} {({d_{colour} \over d_{mag}})}^2 + {\sigma^2(colour)}.
\label{E15}
\end{equation}

To determine the first component on the right hand side of Eq.~(\ref{E15}), we assume $\sigma (V_{HB})=0.03\ \mbox{mag}$ and calculate the change in colour of the fiducial point when its luminosity is shifted by that amount. For the latter term we calculate the median $\sigma(colour)$ at the fiducial point luminosity. The quantity $\sigma(\Delta colour)$ thus depends mainly on the photometric quality in each filter and to some extent on the slope of the MS at the fiducial luminosity. 
The uncertainty of $R_V$ for each filter combination is then determined by the propagation of those errors from $(\sigma(\Delta colour))$ to $\sigma(R_V)$. Here, the governing criterion is the sensitivity of each filter combination to $R_V$, which is discussed in the next section. 
Finally, each filter combination yields an individual best fitting value for $R_V$ together with an interval where the colour excess is reproduced to within its observational uncertainty limits. 
In addition to the observed random errors discussed here, the main sources of systematic errors are discussed in section \ref{dusttype_sourcesofsystematicuncertainties}.

\subsubsection{Independent Filter Combinations}
Not all filter combinations have the same uncertainty and not all of them yield independent results. We therefore have to decide which colours we want to use to minimize the final value of $\sigma(R_V)$.

\begin{figure}[t]
\begin{center}
\includegraphics[width=0.5\textwidth]{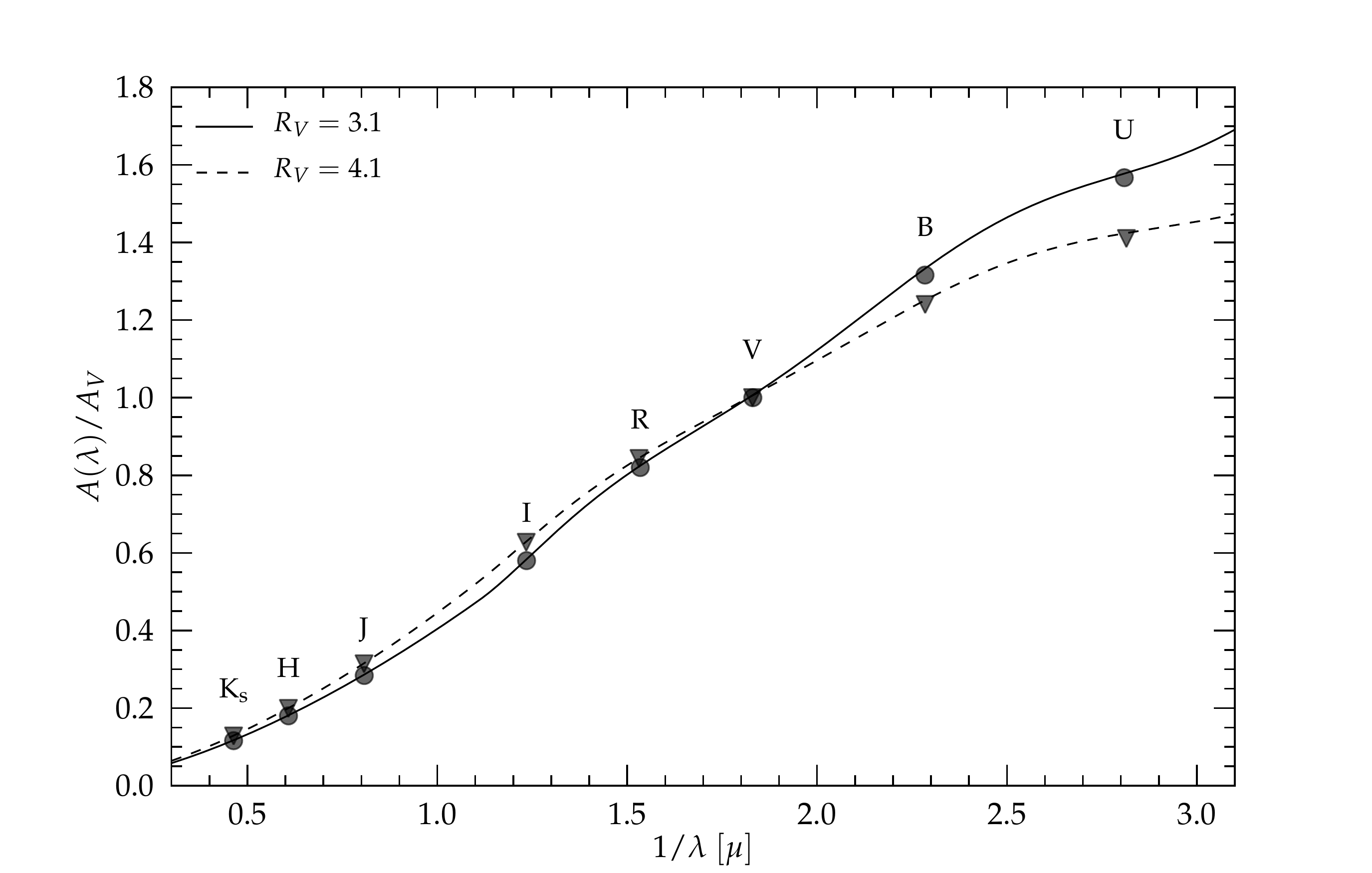}
\caption{Reddening law for different values of $R_V$. The effective wavelength of our filter set is indicated with circles and triangles for $R_V=3.1$ and $R_V=4.1$ respectively. The change of the reddening law with the dust type parameter is the crucial key to determine the dust type relevant to M4.} 
\label{cardelli law}
\end{center}
\end{figure}

The first decision which has to be made is the reference colour to which other combinations shall be compared, where the crucial criterion is to maximize the sensitivity of the ratio $E(\Lambda_1-\Lambda_2)/E(\Lambda_{ref1}-\Lambda_{ref2})$ to $R_V$. This sensitivity is defined by the \textit{difference} in the effect for the correction factor of a given filter combination compared to the correction factor of the reference colour when $R_V$ is changed. Figure~\ref{cardelli law} illustrates the effect of $R_V$ on our filters. It shows the relative extinction as a function of wavelength, assuming two different values of $R_V$.
Regarding this figure, the most promising reference colour is $E(B-V)$, because its change with $R_V$ \textit{differs the most} from that of all other combinations.
From this perspective, $E(U-V)$ would be an even better choice. However, with an effective wavelength of $\sim 3\,560$~\AA, the $U$ filter is located on the extreme short-wavelength end both of common detector sensitivities and of the atmospheric transmission. It is therefore very sensitive to the exact atmospheric conditions and the instrument response. Moreover, the bandpass falls on the Balmer convergence and jump in stellar spectra, which makes the filter sensitive to even small variations in $T_{\mathrm{eff}}$ and $\log g$. 
Consequently, we expect significant uncertainties in the synthetic $U$ magnitudes which would cause a systematic error in the determination of $R_V$.   

The uncertainty of every individual comparison colour depends mainly on two factors: the photometric quality of the data in the given filters and the sensitivity of the ratio ${E({\Lambda_1-\Lambda_2})}/{E(B-V)}$ with $R_V$. The latter factor is optimized by the filter combination spanning the highest wavelength range and the biggest difference from $E(B-V)$ regarding $R_V$-sensitivity.
Given $E(B-V)$ as the reference value, we can assign a maximum of four independent colours contributing to the determination of the dust type under the assumption that results which include the $U$ filter are not trustworthy and that $R$ can be reliably predicted from $V$ and $I$, so it does not really provide independent information.

To minimize systematic errors within the individual bandpasses, we choose the colours in a way such that each filter appears on alternating sides. Therefore we use $E(V-I)$, $E(I-J)$, $E(J-K)$, and $E(B-K)$ as independent colour measurements to be compared to $E(B-V)$. This choice might seem surprising at first, since those filter combinations all have relatively large individual $\sigma(R_V)$ compared to $E(V-J)$ or $E(V-K)$ for example (see Table~\ref{uncertainties1}). However, by using alternating combinations we are reducing any systematic errors from faulty colour-temperature relations (CT-relations) used by the isochrones or bad zero-point calibrations of the observed data to the standard system; the resultant high accuracy outweighs the high precision of some individual colours.
The method therefore necessarily relies on a combination of optical and NIR photometry, since otherwise only $E(V-I)$ could confidently be used to determine the dust type, which would dramatically increase both the systematic and statistical errors in the result.  

\subsection{Isochrones}
\label{dusttype_isochrones}
The method we describe here to determine the dust type toward a stellar population relies fundamentally on the assumption that theoretical isochrones fit observational data either if the correct reddening law is used, or---of course---when there is no reddening. The main source of uncertainty in the fit of isochrone models to observed fiducial sequences comes from CT-relations, which predict a synthetic colour from theoretical surface temperatures and gravities (from model atmospheres) as well as from the predicted $T_{\mathrm{eff}}$-scales, while the theoretical luminosities are considered to be more robust (\citealt{VandenBerg05}). 

We use the newest generation of Victoria-Regina isochrones to determine the colour offsets in different filter combinations. 
These models have been significantly updated since the latest description in \citet{VandenBerg_06}. A summary of recent changes is provided by \citet{Broogard11}, but the most important is the treatment of the gravitational settling of helium (see \citealt{Proffitt91}).
Besides theoretical sequences reaching from the MS up to the RGB tip, we use ZAHB models to determine $(m-M)_V$, which is necessary to match our fiducial point luminosity between observed CMD and isochrones. Once the reddening and the dust type are known, we can further use the HB as a standard candle to determine the distance to the cluster. 

CT-relations and the overall fit of Victoria-Regina isochrones have been tested intensively for both optical and near-infrared filters in two recent studies:
for optical bands \citet{VandenBerg10} found very good agreement between Victoria-Regina isochrones and data for local Population II subdwarfs with Hipparcos-based distances. In a further test, the isochrones reproduced the observed fiducial sequences for several globular clusters with different metallicities very well. For their study, they tested three different CT-relations: the empirical transformations by \citet{Casagrande10}, those derived from MARCS model atmospheres and the semi-empirical relations by \citet{VandenBerg03}, for which they found only minor differences. According to this paper, the models fail to match the observed sequences only for faint MS stars with $M_V\geq6.5$.

\citet{Brasseur10} tested Victoria-Regina isochrones using MARCS model atmosphere-based CT-relations for the 2MASS $J$, $H$ and $K_s$ bands against the same sample of field subdwarfs and found good agreement between the models and the data. Only a small offset was detected in $J-K$ and $V-K$, where the models seem to predict those colours too red by $\sim 0.03$ mag each. Comparing isochrone sequences to the observed CMDs of globular clusters generally yields good agreement for all except the lowest metallicity systems ($\mbox{\hbox{\rm [Fe/H]}} \leq-2.0$). Those discrepancies, however, are limited to the lower RGB segments, where the isochrones seem to be significantly too red, whereas the location of the MS matches for all metallicities.

With the point on the MS 5.5 mag fainter than the ZAHB and a metallicity in M4 of $\mbox{[Fe/H]}=-1.0$, we satisfy both critical conditions for magnitude and metallicity very well, so that we are confident that the models should match the data if the correct reddening law is adopted. 

\begin{figure}[t]
\begin{center}
\includegraphics[width=0.5\textwidth]{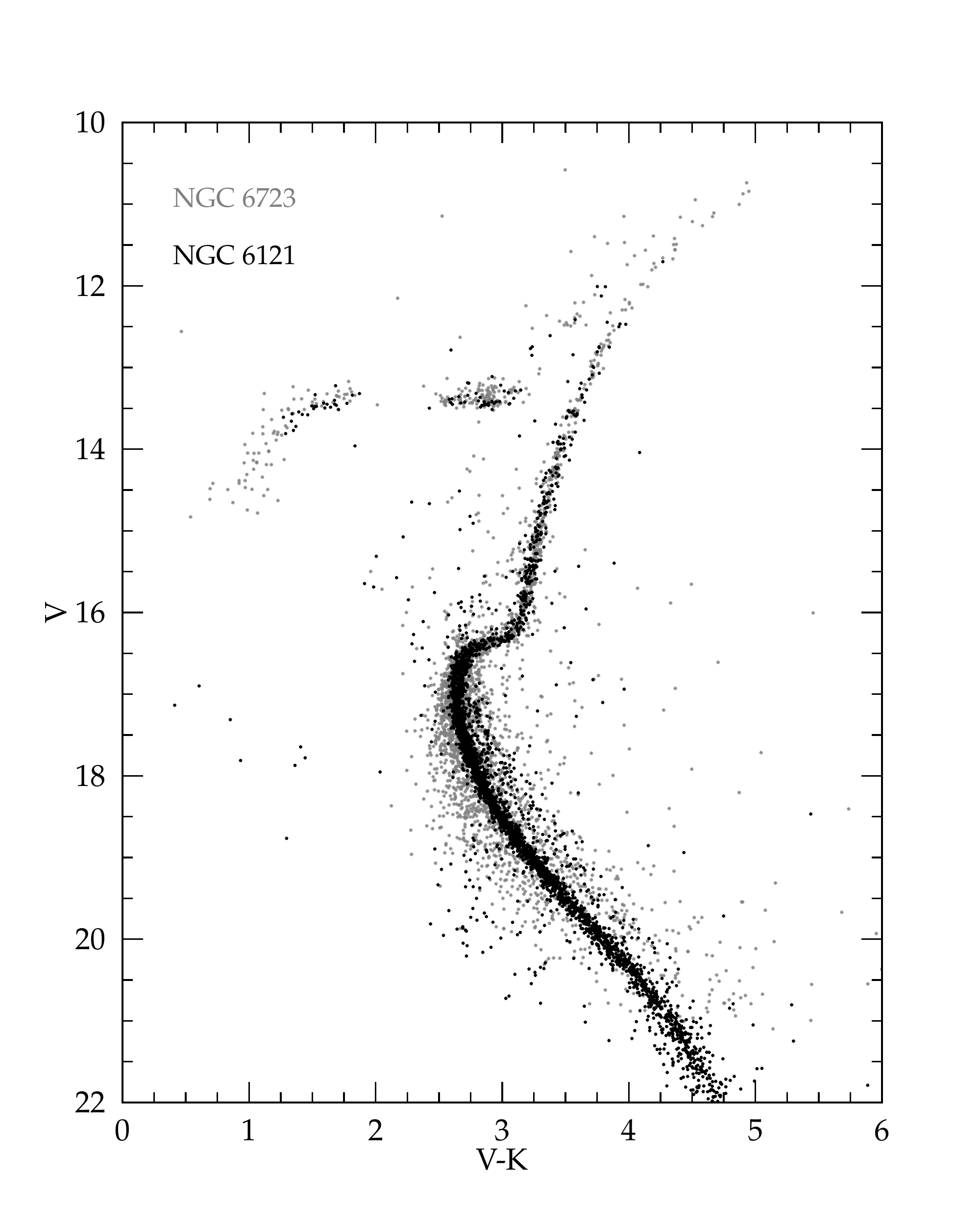}
\caption[Direct comparison between the fiducial sequences of M4 and NGC 6723.]{Direct overlay of M4 (black) and NGC 6723 (grey). The data for NGC 6723 have been shifted by $\Delta (V-K)=+1.07$ and $\Delta V=-2.06$ to match the sequence of M4 (NGC 6121). Although the MS for NGC 6723 is too poorly defined to compare the sequences directly at large magnitudes, both clusters show a remarkable consistency in their RGBs, SGBs, and HB sequences, implying that both systems are almost identical in metallicity and age.} 
\label{overlay}
\end{center}
\end{figure}

\subsubsection{The Guinea Pig: NGC 6723}
We use optical and NIR photometry for the nearly unreddened globular cluster NGC 6723 ($E(B-V)\approx0.05$ according to \citealt{Harris10}) to test the morphological fit and the consistency of our isochrones in different observed colour combinations under optimal conditions. NGC 6723 is almost identical to M4 in chemistry and age and is therefore perfectly suited as effectively unreddened reference cluster. In Figure~\ref{overlay} we correct NGC 6723 to the distance modulus and reddening of M4 to obtain a direct overlay of the two systems. Both sequences show exactly the same slope along the RGB and continuing to below the TO, which implies almost identical ages and chemical abundances. On the lower MS, the photometric quality of NGC 6723 is too poor to allow a precise comparison. Remarkably, even the HB sequences are superimposed precisely with a similar bimodal distribution of red and blue HB stars---which is unusual, if one considers their relatively high metallicities.

\begin{figure}[t]
\begin{center}
\includegraphics[width=0.5\textwidth]{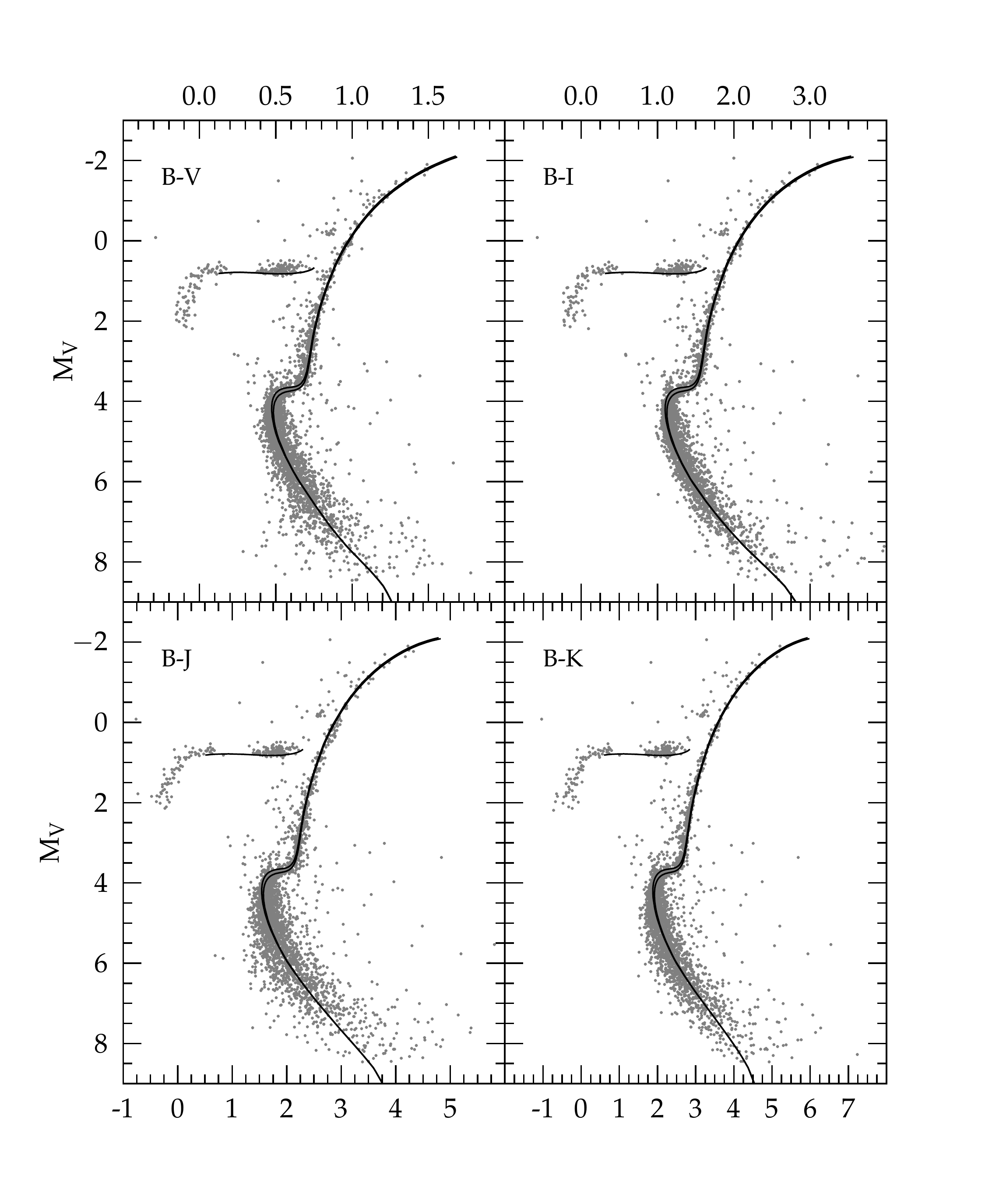}
\caption[Isochrone fits for different filter combinations of NGC 6723.]{Fits of 11 and 12~Gyr isochrones for $Y=0.25$, $\mbox{[Fe/H]}=-1.0$, and $\mathrm{[}\alpha\mathrm{/Fe]}=+0.4$ to various filter combinations of NGC 6723. Using a reddening of $E(B-V)=0.05$ and a standard reddening law with $R_V=3.1$, the isochrones show a good match to the observations for each combination.}
\label{ngc6723_4pack}
\end{center}
\end{figure}

For NGC 6723, a metallicity of $\mbox{\hbox{\rm [Fe/H]}}=-1.0$, an age of 11--12~Gyrs and an assumed reddening of $E(B-V)=0.05$ yield the best fit for our isochrones and are in good agreement with the latest edition of the Harris catalog (\citealt{Harris10}) which gives $\mbox{\hbox{\rm [Fe/H]}}=-1.1$ and $E(B-V)=0.05$. We further adopt [$\alpha\mbox{/Fe]}=+0.4$ following the observational measurements from \citet{Fullton96}, who found $+0.42$ for the alpha elements in this cluster. 

The result is shown in Figure \ref{ngc6723_4pack} for the colours $B-V$, $B-I$, $B-J$, and $B-K$. With an adopted reddening of $E(B-V)=0.05$ and a standard reddening law with $R_V=3.1$ we find an excellent match between the observed CMDs and models for all filter combinations. This test demonstrates that, at least in the intermediate metallicity regime, our isochrones fit data consistently well in all filter combinations in the case where there is no significant reddening. 

\subsubsection{Isochrone Parameters for M4}
Incorrect assumptions for the isochrone parameters [Fe/H], [$\alpha$/Fe] or helium mass fraction ($Y$) will inevitably lead to systematic errors in the determination of $R_V$. For M4 we are using isochrones with $\mathrm{[Fe/H]}=-1.0$, $\mathrm{[}\alpha\mathrm{/Fe]}=+0.4$, $Y=0.25$ and an age of 12\,Gyrs which we will now justify. 

\begin{figure}[t]
\begin{center}
\includegraphics[width=0.5\textwidth]{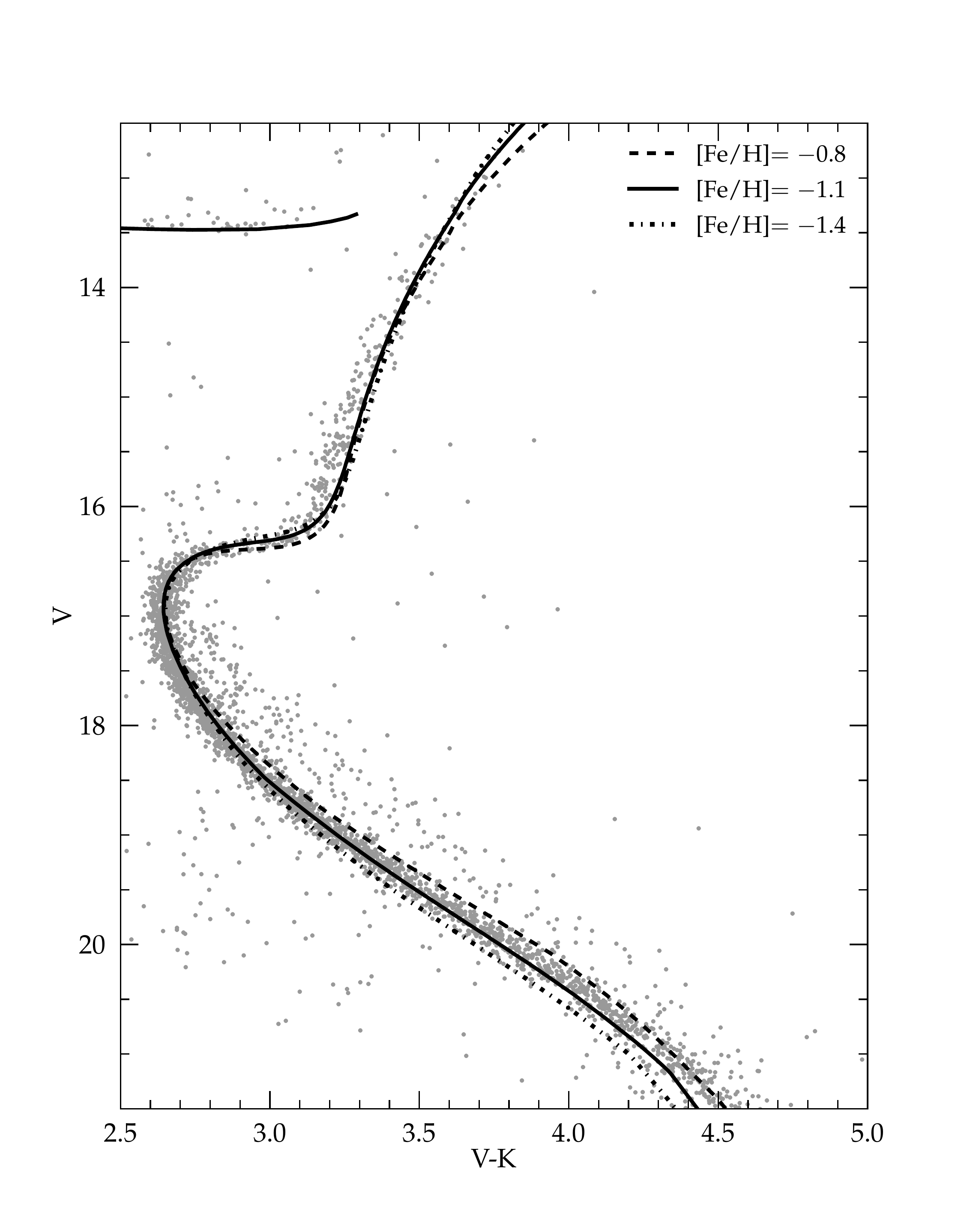}
\caption[Estimate of the M4 metallicity from isochrone fits.]{The overall shape of the the observed fiducial sequence is used to constrain the metallicity of M4 by isochrone fitting. Using 12~Gyr isochrones, we find a best fit when a value of $\mathrm{[Fe/H]}\approx -1.0$ is adopted.}
\label{determine metal}
\end{center}
\end{figure}

Although the main-sequence fiducial point which we are using to determine the colour excess is independent of age, it is important to know the metallicity of the cluster as accurately as possible since this parameter affects the shape and location of the isochrone, the absolute magnitude of the ZAHB, and also the transformation-factor zero points from the reddening law. There are many studies of the metallicity of M4, yielding values between $\mbox{\hbox{\rm [Fe/H]}}\approx-1.4$ and $-1.0$ with recent results tending to favour the metal rich end of this interval. For example, \citet{Mucciarelli11} found $\mbox{\hbox{\rm [Fe/H]}}=-1.1\pm0.07$ for 87 MS and RGB stars in M4 and \citet{Marino08} derived $\mbox{\hbox{\rm [Fe/H]}}=-1.07\pm0.01$ in their spectroscopic study of 105 RGB stars.  
When we fit isochrones to our data for M4, we also find a best fit for a value of [Fe/H] close to $-1.0$. Figure~\ref{determine metal} shows a comparison between isochrone fits with metallicities between $\mbox{\hbox{\rm [Fe/H]}}=-1.4$ and $-0.8$, and shows that the overall slope of the observed fiducial sequence is matched best with $\hbox{\rm [Fe/H]}=-1.0$ and an age of $12~\mbox{Gyrs}$.
The recent work of \citet{Bedin09} and \citet{Mucciarelli11} supports this parameter choice: they both find a best fit for isochrones with $\mbox{\hbox{\rm [Fe/H]}}=-1.01$ and age of 12~Gyrs derived from the WD cooling sequence. This latter number is in good agreement with the results from \citet{Hansen04} who find a best fit for $12.1~\mbox{Gyrs}$ with the same method, which gives us confidence that our choice of isochrone parameters does not add a significant systematic error to the derived value of $R_V$.
Since metallicity seems to be the parameter with the highest uncertainty and the biggest impact on $R_V$, we derive a quantitative estimation of $\sigma(R_V)$ due to uncertain [Fe/H] assumptions in section \ref{dusttype_sourcesofsystematicuncertainties}.

For M4, recent alpha-element measurements from \citet{Marino08} yield $\mbox{[Mg/Fe]}=+0.50$, $\mbox{[Ca/Fe]}=+0.28$, $\mbox{[Si/Fe]}=+0.48$ and $\mbox{[Ti/Fe]}=+0.32$ and therefore confirm the generally enhanced alpha-elements in old, metal-poor GC stars and justify our choice of [$\alpha\mbox{/Fe]} = 0.40$.

GC helium abundances are (still) assumed to be similar or equal to the primordial helium abundance arising from Big Bang nucleosynthesis. This would imply a helium abundance near a value of $Y=0.25$ (\citealt{Cyburt08}), at least for the vast majority of clusters. There is no evidence in the literature that M4 is an exception to that rule. 

\subsection{Results}
\label{dusttype_results}
For the colour excess in $B-V$ we find $E(B-V)=0.37\pm0.01$.
Every colour offset we are using in direct comparison to $E(B-V)$ yields a best-fitting value together with an acceptable uncertainty interval for $R_V$ that produces the necessary observed excess for this filter combination. From those individual results we derive the final value of $R_V$ with the independent sample to gain the smallest possible random and systematic errors.

\subsubsection{Individual Filter Combinations and the Independent Sample}
\begin{figure}[t]
\begin{center}
\includegraphics[width=0.5\textwidth]{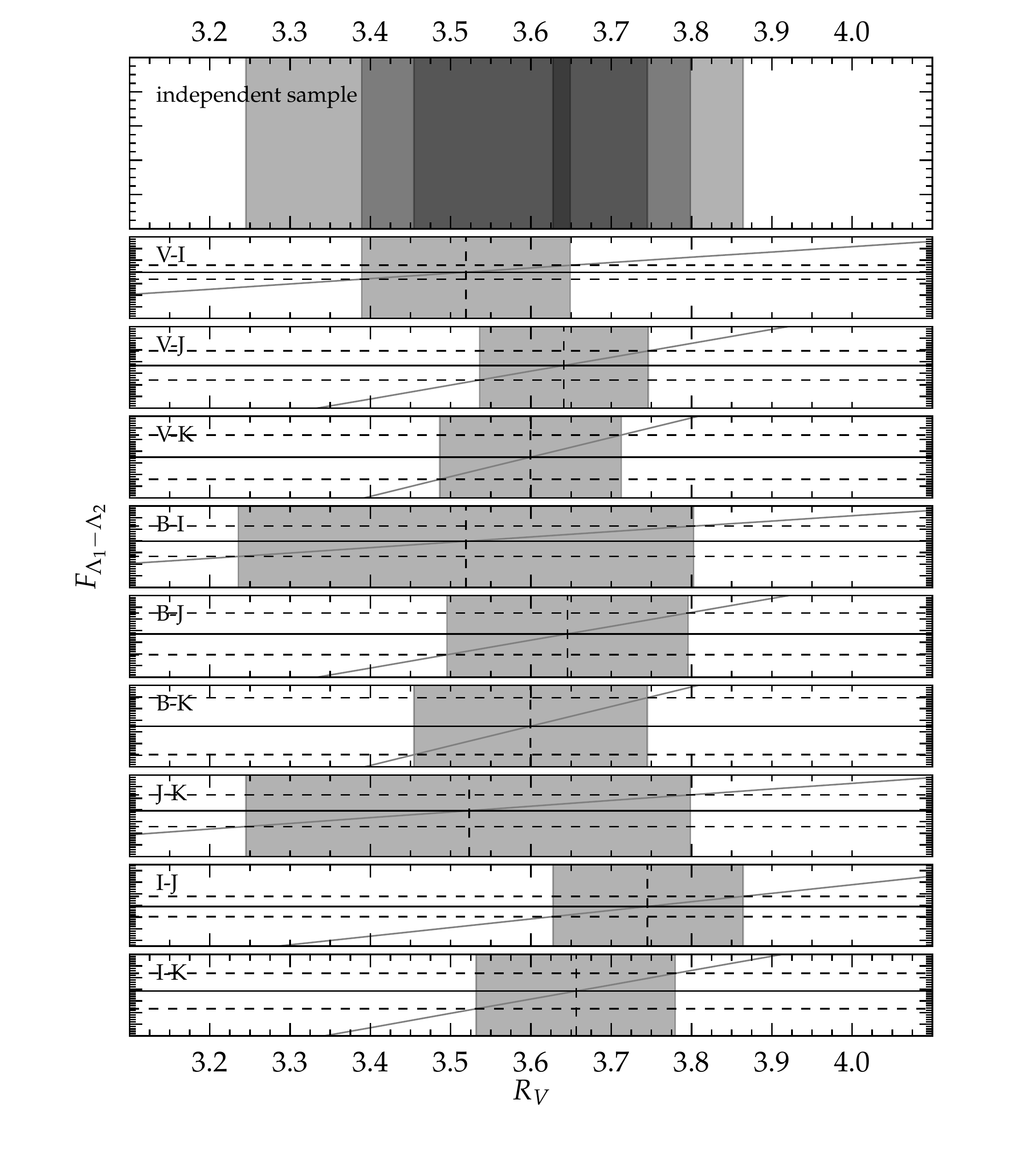}
\caption[Determination of the dust-type parameter $R_V$ in different filter combinations.]{Visualization of the determination of the dust type in the line of sight to M4: grey lines represent the dependence of $F_{\Lambda_1-\Lambda_2}$ on $R_V$. Solid black horizontal lines indicate the best fitting value of $F_{\Lambda_1-\Lambda_2}$ for each filter combination from observations and dotted black lines mark the uncertainty interval. Each filter combination produces a shaded region where the boundaries are defined by the intersection of the dotted black lines with the corresponding grey line. In the top panel, the results for the four independent combinations $V-I$, $I-J$, $J-K$ and $B-K$ are over-plotted in a way that overlapping areas produce a more saturated shade.}
\label{dusttype2} 
\end{center}
\end{figure}

In a first step, we want to find out whether, in general, it is possible to reach a consistent fit for all colours within their photometric uncertainty limits. For that purpose we superimpose the individual results by shading the acceptable $R_V$ intervals for each colour with the same intensity and in such a way that overlapping areas produce stronger shades. The top panel of Figure~\ref{dusttype2} shows such an overlay for the independent colours $V-I$, $I-J$, $J-K$ and $B-K$. Using this colour selection, we find best fitting values in the range of $R_V=3.52 - 3.75$ with only a small interval where the model matches the data consistently for \textit{all} colours at $3.62\leq R_V \leq 3.65$. 

The rather symmetric distribution around this best fitting region from colours where the filters alternate signs suggests that the deviations between different colours are mainly caused by systematic errors in individual filters. Similarly, it demonstrates that those errors are small compared to the observational uncertainty, since a region of common overlap still exists.

\begin{table}[t]
\centering
\caption{Results for the individual filter combinations.}
\begin{tabular}{cccccccccc}
\hline
\hline

$\Lambda_1 - \Lambda_2$& Offset			& $F_{\Lambda_1-\Lambda_2}$	& $R_V$				 	\\
(1) & (2) & (3) & (4) \\\hline 
   $B-V$	   	& $0.37 \pm0.01$	 		& $1.00$			&      $-$							\\
   $V-I$          	& $0.53 \pm0.01$ 			& $1.45 \pm0.03$	&	$3.52 \pm0.13$				\\
   $V-J$         	&$0.97 \pm0.02$			& $2.63 \pm0.06$	&	$3.64 \pm0.11$				\\
   $V-K$ 		&$1.21 \pm0.04$			& $3.28 \pm0.10$	&	$3.60 \pm0.11$				\\	
   $B-I$		&$0.90 \pm0.02$			& $2.45 \pm0.07$	&	$3.52 \pm0.28$				\\
   $B-J$		&$1.34 \pm0.03$	       		& $3.64 \pm0.09$	&	$3.65 \pm0.15$				\\
   $B-K$		&$1.57 \pm0.05$			& $4.28 \pm0.12$	&	$3.60 \pm0.14$				\\
   $J-K$		&$0.24 \pm0.03$			& $0.65 \pm0.07$	&	$3.52 \pm0.28$				\\
   $I-J$   		&$0.44 \pm0.02$ 		        	& $1.20 \pm0.04$  	&	$3.75 \pm0.12$				\\
   $I-K$         	&$0.68 \pm0.03$    	 		& $1.84 \pm0.08$  	&	$3.66 \pm0.13$				\\\hline

\end{tabular}
\tablecomments{(2): Empirical colour offsets together with observational uncertainties between our data and a theoretical isochrone with $\mbox{\hbox{\rm [Fe/H]}}=-1.0$, [$\alpha$/Fe]$=+0.4$ and an age of 12~Gyrs. In (3) are the resulting transformation factors. (4) gives the individual best-fitting dust-type parameters for the line of sight to M4. The uncertainties in (3) and (4) represent the propagation of the error intervals from (2).}
\label{uncertainties1}  
\end{table}

Next, we find the best fitting dust type for M4 by evaluating the four selected colour combinations mentioned above, assuming that each of them yields an independent measurement of $R_V$. We determine the best fitting value of $R_V$ and its uncertainty with the weighted mean and standard deviation of the results for the different filter combinations and find $R_V=3.62 \pm0.07.$ This result is in good agreement with previous estimates in the literature, but with a total uncertainty several times smaller.

\begin{figure}[]
\begin{center}
\includegraphics[width=0.5\textwidth]{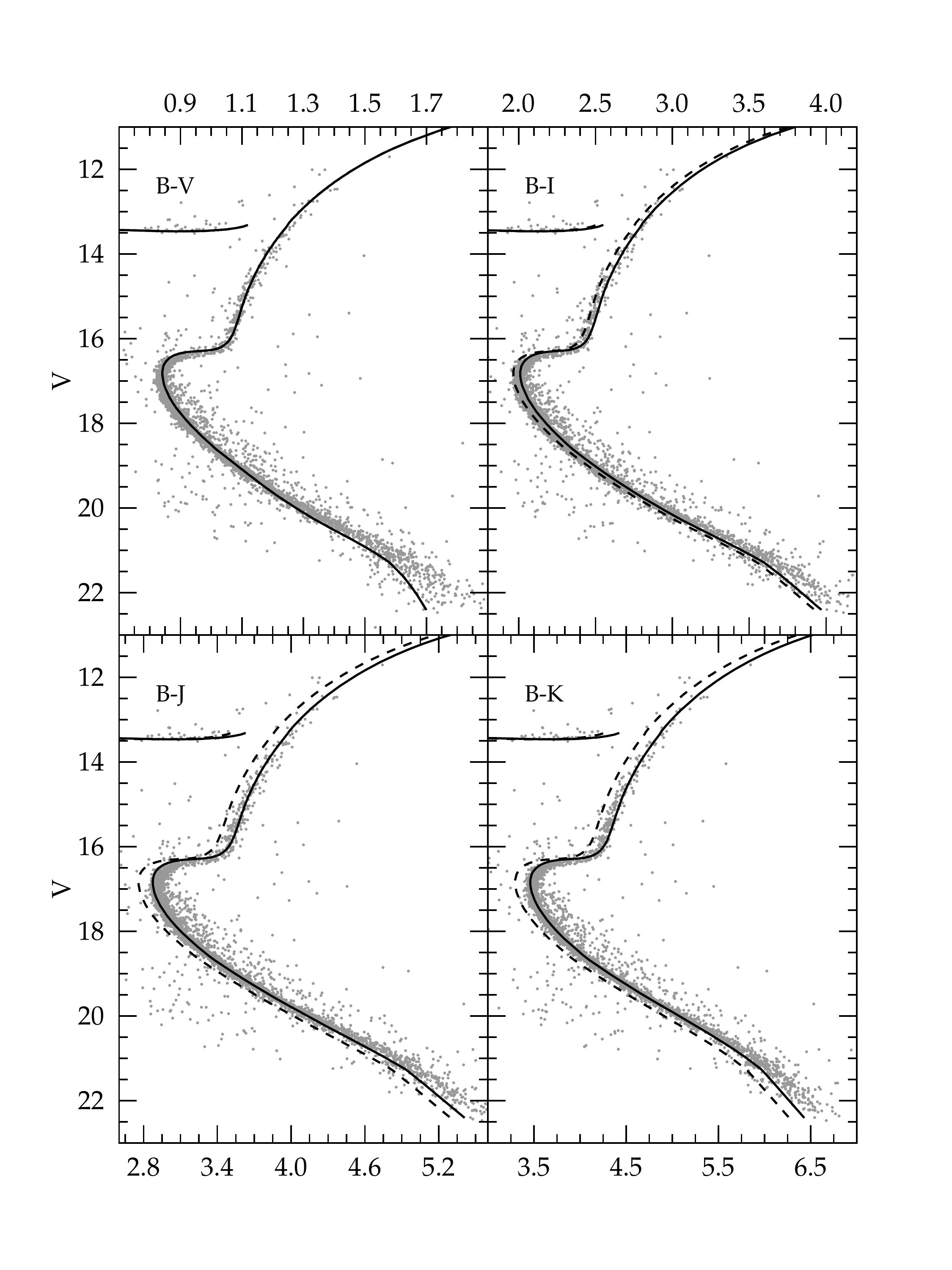}
\caption{Direct comparison of two different laws: In each panel, the dashed line shows an isochrone corrected with a standard reddening law ($R_V=3.1$), while a solid line shows an isochrone if a reddening law with $R_V=3.62$ is used instead. Since both corrections are normalized to $B-V$, both isochrones have identical loci here.}

\label{final comparison} 
\end{center}
\end{figure}

The individual best fitting $R_V$ values for the various colours as well as the measured colour offsets and the values of $F_{\Lambda_1-\Lambda_2}$ that are needed to produce these offsets are listed in Table~\ref{uncertainties1}, together with their respective uncertainties. Note that the value derived as the weighted mean nearly reproduces the region of total overlap. Furthermore, all filter combinations deduced from the independent sample yield best fitting values close to and symmetrically distributed around this mean. In Figure~\ref{final comparison} we demonstrate the improvement in the isochrone fits when we use our derived value of $R_V$ instead of a standard law with $R_V=3.1$ and Table~\ref{rv in literature} we compare our result for $R_V$ and its uncertainty to other results obtained in the past. Note that indirect measurements (i.e., along nearby lines of sight) can give a general hint concerning the dust type of M4, but cannot actually be used for an accurate prediction.

It is important to point out that the dust- type parameter $R_V$ determined here is not exactly the same as the actual total-to-selective extinction ratio $A_{V}/(A_{B}-A_{V})$ for our filters. This is due to the fact that the dust type parameter $R_V$ in the Cardelli equations is defined for Johnson's $B$ and $V$ bands using O and B type stars. As soon as a different filter set is used, $R_V$ describes only the shape of the reddening law and not the extinction ratio. The same is true if the reddening law is used for stars with significantly different temperatures, as is the case in our study. The difference is caused by the change in effective wavelengths for the $B$ and $V$ filters which is, in a sense, equivalent to using a different filter set.

For the reddening correction and the distance determination, however, we must use the actual ratio between $A_{V}$ and $E(B-V)$. The extinction values we obtain with a reddening law using $A_V=1.317$ and $R_V=3.62$ as applied to a star with the atmospheric parameters of our main-sequence fiducial point are listed in Table~\ref{correct extinctions} and yield $A_{V} / (A_{B}-A_{V})=3.76 \pm 0.07$ where the confidence interval represents the uncertainty of the total-to-selective extinction ratio for a reddening law appropriate to {\it these\/} stars observed with {\it this\/} equipment.
\newline
\

\begin{table}[h]
\centering
\caption{Comparison of the determined value of $R_V$ to literature values.}
\begin{tabular}{lccccccccc}
\hline
\hline

Author					& Quoted $R_V$	& $\sigma$		& Method  	& Notes			 	\\\hline 
   Cudworth \& Rees (1991)	& 3.3	 		& 0.7			&      direct	& 					\\
   Dixon \& Longmore (1993)	& $\sim 4$		& -				&	direct	&					\\
   Peterson et al. (1995)		& $\sim 4$		& -				&	direct	&					\\
   Ivans et al. (1999)			&3.4				& 0.4			&	direct	&					\\	
   Chini (1981)		  		&4.2				& 0.2			&	indirect	&  (3)  				\\
   Clayton \& Cardelli	 (1988)	&3.8	       			& -				&	indirect	&  (2), (3)				\\
   Vrba (1993)				&3.99			& 0.18			&	indirect	&  (1), (3)				\\
   This Work				&3.62			& 0.07			&	direct	&					\\\hline

\end{tabular}
\tablecomments{(1): This result was obtained by the measurement of 12 stars in the Sco-Oph cloud complex. The stated value (that is used by \citet{Richer97} and others) only represents the mean value of all measurements, where the individual results actually vary between $R_V=3.15$ and $R_V=5.25$(!). This study is an excellent example of the fact that it is not possible to infer the value of $R_V$ for an object from its direct environment. Specifically, even if the lines of sight are similar, we cannot evaluate possible inhomogeneities transverse to the line of sight, or along the line of sight between the nearer and more distant objects.
(2): This result is obtained from the measurement of only one star. 
(3) The value actually represents the observed ratio $A_{V}/(A_{B}-A_{V})$ rather than the dust-type parameter for the Cardelli et~al. reddening law.}
\label{rv in literature}    
\end{table}

\begin{table}[t]
\begin{center}
\caption{Best fitting filter extinctions for M4}
\begin{tabular}{cccccccccc}
\hline
\hline
$R_V$	&   $A_{U}\over A_{V}$	&   $A_{B}\over A_{V}$	&   $A_{R}\over A_{V}$	&   $A_{I}\over A_{V}$	&   $A_{J}\over A_{V}$	&   $A_{H}\over A_{V}$	&   $A_{K}\over A_{V}$  \\\hline
       $3.62$& 1.474& 1.266& 0.831& 0.608& 0.302& 0.191& 0.123\\\hline    

\end{tabular}
\label{correct extinctions}
\end{center}
\end{table}

\subsubsection{A Quick Test}
The two colours $V-J$ and $B-I$ provide a special case since the values of their transformation factors cross near the canonical dust type at $R_V\approx3.2$ with $E(V-J)$ being smaller for $R_V \leq 3.2$ and $E(B-I)$ being smaller for $R_V > 3.2$. Additionally, both filter combinations have about the same theoretical colour around the MSTO for all metallicities below $\mbox{[Fe/H]}\approx-0.8$. Therefore those colours provide a quick qualitative test for the dust type: if both $B-I$ and $V-J$ have about the same colour around the MSTO in an observed CMD, a standard dust type can be assumed. If $V-J$ is significantly redder than $B-I$, $R_V > 3.2$; if $V-J$ is significantly bluer, $R_V < 3.2$. 

\subsection{Sources of Systematic Uncertainties}
\label{dusttype_sourcesofsystematicuncertainties}
Since we are using observational data together with model assumptions, there are several different sources of uncertainty contributing to the total uncertainty of $R_V$. After we estimated the random error in section \ref{dusttype_error_estimation}, we now discuss the different possible sources of systematic uncertainty and their significance.

The assumption of [Fe/H] affects the position and shape of the isochrone, the magnitude of the ZAHB and also the zero points for the reddening law. Since the uncertainty in metallicity appears to have the biggest impact on the accuracy of our derived $R_V$ and consequently on the derived distance to M4, we estimate the change in $R_V$ arising from an uncertainty in [Fe/H], by performing the whole procedure to derive $R_V$ while varying [Fe/H] by $\pm0.2$~dex. This includes the use of a different isochrone to determine the ZAHB luminosity and the colour offsets as well as a different set of zero points for the transformation factors in our reddening law.
The result for $\mbox{\hbox{\rm [Fe/H]}}=-1.2$ yields a best-fitting value for $R_V=3.56 \pm0.07$ with a predicted reddening of $E(B-V)=0.39$. An assumption of $\mbox{\hbox{\rm [Fe/H]}}=-0.8$ yields $R_V=3.96 \pm0.09$ and $E(B-V)=0.30$.
By comparing these numbers to the result we got assuming $\hbox{\rm [Fe/H]}=-1.0$, the systematic effect on $R_V$ due to uncertainties in [Fe/H] is highly asymmetrical and shows much larger deviations towards higher metallicities. However, in our case, the high-metallicity scenario can clearly be ruled out not only from isochrone fitting and from the unreasonably low amount of reddening resulting from this assumption, but also from recent spectroscopic studies.  
Therefore, we allow only the lower metallicity as a possible scenario and conclude that an uncertainty of $\sigma\mbox{\hbox{\rm [Fe/H]}}=\pm0.2$~dex yields a change of $R_V\approx \pm0.06$. 

The value of $R_V$ that best fits the data further depends upon the exact reddening law we are using, and explicitly on the zero points for the transformation factors $F_{\Lambda_1-\Lambda_2}$. These values differ between different literature sources and our own calculations, and the same would be the case for the value of $R_V$ needed to match isochrones with observations. We can only estimate the uncertainty introduced by the definition of the reddening law zero points by assessing the change in $R_V$ when we use different sets of values from the literature and our own. The effect on $R_V$ is, in any case, smaller than $0.1$ which might define an upper limit for the uncertainty due to the definition of the reddening law. However, most of the discrepancy here is introduced by different assumptions for the stellar atmospheric conditions. Once those are assumed correctly, the actual uncertainty arising from the zero points is only caused by the quality of the spectra and might be significantly smaller than the number stated above.

When we determine observational colour excesses in different filter combinations relative to theoretical isochrones, we assume that the models actually predict the unreddened CMD colours correctly. If this is not the case, for example due to faulty CT-relations or $T_{\mathrm{eff}}$-scales, the result of $R_V$ is biased. We discussed this issue in section \ref{dusttype_isochrones} where we found that the isochrones provide a consistent fit to the nearly unreddened cluster NGC 6723. However, minor discrepancies for at least some individual filters are likely to exist as our results for the individual filter combinations show.

The observed data are calibrated to the \citet{Landolt92} standard system for Johnson-Cousins $UBV(RI)_c$ and to the 2MASS standard system for the NIR counterparts $J$ and $K_s$. 
The accuracy of the match between our adopted instrumental bandpasses and those from the respective standard systems is determined by the level to which any systematic differences between both can be corrected. Inaccuracies in the calibration to the standard system have an impact similar to that of faulty CT-relations as they systematically bias the colour offset between model and data. 
Generally, the calibration for our data shows no systematic deviation from this standard system. However, a final intrinsic uncertainty of as much as $\sigma (colour) \approx \pm0.03$ (see discussion in \citealt{Stetson05}) has to be considered, especially if the data were obtained in just one observing run, as is the case for our NIR photometry.
The calibration of the optical photometry of M4 was derived independently on 20 photometric nights distributed among 9 observing runs, so the overall standard calibration should be significantly better than 0.01~mag in this case.

By using alternating colours such as $V-I$, $I-J$, $J-K$, and $B-K$ to determine the value of $R_V$, we largely eliminate any systematic error which is specific for the individual filters $I$ and $J$. For this reason, the uncertainties in CT-relations and the calibration to the standard system have only a minor impact on the quality of our result. However, the uncertainty in our knowledge of [Fe/H] and the exact reddening law zero point is not filter-specific and therefore has to be taken into account when the total uncertainty of $R_V$ is estimated.  

\subsection{Implications for the Distance Modulus of M4}
The absolute distance to a globular cluster, or its true distance modulus $DM_0$, strongly depends on the reddening law and the assumed dust type, if it is derived from standard candles of a stellar population such as variable stars or the apparent magnitude of the ZAHB. In this case 
\begin{equation}
DM_{0}= DM_{\mathrm{app}} - A_{\Lambda}
\end{equation}
and
\begin{equation}
d[kpc]={{10^{({{[DM_{0}}/{5}]}+1)}} \over {1\,000}}
\label{E_distance}
\end{equation}
where $DM_{\mathrm{app}}$ is the observed or apparent distance modulus and $A_{\Lambda}$ the extinction correction for the filter $\Lambda$. However, the extinction is usually determined with the colour excess and expressed by $E(B-V)$. $A_{\Lambda}$ is then obtained with the ratio of absolute to relative extinction $R_{\Lambda}$:
\begin{equation}
A_{\Lambda}=R_{\Lambda} E(B-V).
\end{equation}
Each $A_{\Lambda}$ is therefore unambiguously correlated with the dust type, as is $R_V$, and as a consequence $DM_0$ can only be determined as a function of that parameter.

We determine the apparent distance modulus by comparing the ZAHB $V$-band magnitude to models and find
$(m-M)_{V}=12.66 \pm0.03$,
which is slightly smaller than most of the values we find in the literature, especially the value given in \citet{Harris10} of $(m-M)_{V}=12.82$. This is partly because we are using the findings from the treatment of spatial differential reddening in this cluster, which defines a slightly brighter zero-age horizontal branch than the initial, uncorrected data. The use of a slightly higher metallicity for the isochrones than assumed in Harris additionally decreases the distance modulus. 

Using a reddening law with $R_V=3.62\pm0.09$ (including $\sigma$[Fe/H]) and a reddening of $E(B-V)=0.37\pm0.01$ determined with isochrone fitting, we get a total-to-selective-extinction of $A_{V}/(A_{B}-A_{V})=3.76$ for our filters. The distance to M4 is finally calculated with Eq.~(\ref{E_distance}), yielding
$d =1.80 \pm 0.05\,\mbox{kpc}$ or
$(m-M)_{0}=11.28\pm0.06$.

The stated uncertainty for the absolute distance of M4 includes the observational uncertainty for $R_V$ ($\sigma(R_V)=0.09$; including a $\sigma$[Fe/H] of $\pm 0.2$), the observational uncertainty for the apparent distance modulus ($\sigma(DM_{app})=0.03$), and the uncertainty for the reddening ($\sigma(E(B-V)=0.01$).

Some publications already use a higher value of $R_V$, and found a distance to M4 of $d=1.72 \pm0.14\,\mbox{kpc}$ (\citealt{Hansen04}; \citealt{Richer97}). 
Compared to these estimates, our distance is slightly larger but still within their observational uncertainty (see Table \ref{compare}). 

\begin{table}[h]
\centering
\caption{Overview of recently derived absolute distances to M4.}
\begin{tabular}{lccccccccc}
\hline
\hline

Author					&	Distance	                 	& $\sigma$	&	 $A_{V}/E(B-V)$ &  Notes  			 	\\\hline
   This work	   		& 1.80 kpc	 			& 0.05	& 	3.76	     &								\\
   Harris (2010 edition)    	& 2.20 kpc				& -		&	3.20	     &		(1)						\\ 
   Bedin (2009)			& 1.86 kpc				& -		&	3.80	     &    			        				\\		
  Hansen (2004)		& 1.72 kpc				&0.18	&	3.80	     &		(2)						\\
  Peterson (1995)		& 1.72 kpc				&0.14	&	-	     &		(3)						\\\hline

\end{tabular}
\tablecomments{(1): $R_V$ comes from \citet{Cudworth90}, which is the reference for Harris distance estimate. (2): This result is identical to \citet{Richer97}. The uncertainty does not include uncertainties in $R_V$. (3): Derived from astrometric measurements; independent of photometric standard candles.}
\label{compare}    
\end{table}

\section{Summary and Discussion}
\label{chapter_summary}
In this work we have investigated the dust properties in the direction to the globular cluster M4 using near-infrared $J$ and $K_s$ observations obtained with SOFI on the New Technology Telescope in Chile, which have been combined with archival optical Johnson-Cousins $UBV(RI)_c$ photometry. We determined the size and the distribution of the spatial differential reddening across the cluster face, the mean absolute amount of reddening, and finally the type of dust in the line of sight to M4. With this information we were able to determine the absolute distance to M4 with a higher precision than previous estimates.    

We have used colour excesses derived from a combination of NIR and optical photometry together with Victoria-Regina isochrones to determine the dust type parameter represented by $R_V$ in the Cardelli reddening law (\citealt{Cardelli89}) and the actual value of the total-to-selective extinction ratio for M4. Our method is independent of age assumptions, insensitive to metallicity and appears to be significantly more precise and accurate than existing spectroscopic approaches.
In the following we summarize the main results:

\begin{itemize}

\item We investigated the reddening variations across the face of the field of M4, finding a total peak-to-peak difference of $\delta E(B-V)\approx0.2\,\mbox{mag}$ from the north-east to the south-west within a radius of $10^{\prime}$ around the cluster centre. Therefore the differential effects within the field of M4 are about half the size of its mean total reddening, which we found to be $E(B-V)=0.37\pm0.01$. By correcting for the variations in reddening, we have been able to decrease the observational scatter in our photometry by about 50\% and consequently increase the precision of important parameters like the zero-age horizontal branch luminosity by the same factor.   

\item We have written a program to investigate the impact of stellar temperature, surface gravity and metallicity on the extinction properties in different broadband filters and on the necessary reddening corrections. We found similar sized effects for temperature and surface gravity within typical globular cluster parameter limits; each causes a change of about $3\%$ in the necessary correction factor for each filter combination. However, the impact of gravity seems to be significant only for the coolest main sequence and red giant branch stars. It is important to note that metallicity causes similar changes in the correction factors as the other atmospheric parameters when it is changed from solar metallicity ($\mbox{\hbox{\rm [Fe/H]}}=0.0$) to $\mbox{\hbox{\rm [Fe/H]}}=-2.5$. 

\item We made use of three different spectral databases to intercompare the results obtained with synthetic ATLAS9 spectra and observed spectra for different temperatures (using the atlas of \citet{Pickles98}) and metallicities (using the atlas of \citet{Sanchez06}). We found excellent agreement between the different types of spectra for all metallicities and for temperatures above 5\,000\,K; however significant deviations are found at lower temperatures. Inadequate treatment of molecular bands within the ATLAS9 models is suggested to be the most likely cause of the discrepancies at low temperatures.

\item We have used a combination of optical and near-infrared colours to determine the average dust type toward M4, as characterized by $R_V$ in the Cardelli et~al.~reddening law, by fitting the newest generation of Victoria-Regina isochrones to our observed CMD, requiring a consistent fit for all colours. This method is independent of the assumed age of the system and insensitive to the assumed metallicity.
With our method, we have determined the dust-type parameter to be $R_V=3.62 \pm0.07\pm0.06$, where the first uncertainty is the observational precision and the second represents possible systematic effects due primarily to uncertainties in [Fe/H] . Using this dust-type parameter, we find an actual ratio for the total-to-selective extinction in our filters of $A_{V}/(A_{B}-A_{V})=3.76$.
A generally large number for $R_V$ has been proposed several times in the literature but never with such a strong quantitative justification.

\item We have tested our isochrones for optical and NIR colours in the intermediate metal poor regime on the nearly unreddened globular cluster NGC 6723 and confirm that they provide a good and consistent fit for all filter combinations. It is worth noting that NGC 6723 is identical in age and metallicity to M4 within the photometric uncertainties, with a similar bimodal HB sequence. It is therefore an essentially unreddened twin to the highly reddened M4. 

\item With the new dust type and the knowledge about the spatial differential reddening effects, we have reexamined the distance to M4 and find a significantly ($\sim18\%$) smaller value for the absolute distance compared to the parameter provided by \citet{Harris10}. This makes M4 unambiguously the closest globular cluster to the Sun with a distance of $1.80\pm0.05\,\mbox{kpc}$. Our value is, however, somewhat larger than some recent estimates of $d=1.72\,\mbox{kpc}$, based on $R_V=3.8$ (\citealt{Hansen04}).   

\end{itemize}

The proper treatment of interstellar extinction is a crucial step to achieve accurate astrophysical results from photometric observations. As seen from this perspective, M4 is an excellent object to study the effect of interstellar extinction and its impact on the accuracy of the derived parameters due to the strong, differential and unusual extinction it suffers in combination with its extreme proximity, which allows extremely deep and precise photometry in all photometric filters. 
In a way, it serves as an in-situ laboratory from which we can learn about the application of extinction corrections for fainter objects such as stellar populations in the bulge of our Galaxy or resolved extragalactic populations.

Our study reveals that individual---object-specific---parameters of the target of investigations such as temperature, surface gravity and metallicity can all have a significant impact on the extinction properties of broadband filters and consequently on the colour and extinction correction procedure. Even though variations of these parameters \textit{within} the population do not seem to add a significant systematic error for systems with $E(B-V)\leq0.5$, the systematic difference caused by the use of wrong zero points (e.g., defined for a Vega-like star) can be much larger and should be taken into account for all reddened systems. Therefore we recommend the use of an object-specific reddening law defined for appropriate values of the crucial target parameters for the correction procedure. However, For highly reddened targets with $E(B-V)=0.5$ or more, the use of a star-by-star correction should be considered to account for the differential effects within the population.
 
Moreover, our work on M4 has shown that the shape of the reddening law defined by the type of dust in the line of sight can be as object-specific as the stellar parameters. Especially for highly reddened objects, the unconsidered use of the standard dust-type parameter for the interstellar medium---or other generalized reddening laws---can cause significant systematic errors in the interpretation of photometric data.

Finally, we find that model atmospheres and the associated synthetic spectra still have problems adequately predicting the complex properties of molecular bands at cool temperatures. In terms of the derived reddening-correction factors, this effect is small. However, more work needs to be done to investigate the impact of this insufficiency on the actual CT-relations that are used to compare model predictions to observations. 

\acknowledgements
B.~Hendricks gives thanks to Karsten Brogaard for many helpful discussions of several reddening-related issues which are incorporated in this manuscript. This work was supported by the Natural Sciences and Engineering Research Council of Canada through a Discovery Grant to DAV.

\appendix
\section{Fiducial Sequences}
\label{appendix_b}

\begin{table}[H]
\centering
\caption{Empirical fiducial sequence for M4.}
\begin{tabular}{cccccc}
\hline
\hline
$V$   &$(V-J)$	&   $(V-K)$	&  $V_{c}$ & $(V-J)_{c}$	&   $(V-K)_{c}$  \\\hline
12.550& 2.952& 3.922& 12.408& 2.930& 3.884\\
 12.850& 2.840& 3.758& 12.715& 2.819& 3.721\\
 13.150& 2.771& 3.670& 13.014& 2.750& 3.633\\
 13.450& 2.718& 3.605& 13.314& 2.697& 3.568\\
 13.750& 2.658& 3.533& 13.620& 2.638& 3.498\\
 14.050& 2.618& 3.463& 13.925& 2.601& 3.429\\
 14.350& 2.572& 3.397& 14.246& 2.560& 3.369\\
 14.650& 2.537& 3.336& 14.566& 2.527& 3.313\\
 14.950& 2.510& 3.279& 14.870& 2.500& 3.258\\
 15.250& 2.478& 3.244& 15.176& 2.471& 3.224\\
 15.550& 2.459& 3.213& 15.488& 2.453& 3.197\\
 15.850& 2.425& 3.170& 15.777& 2.423& 3.150\\
 16.150& 2.397& 3.126& 16.085& 2.398& 3.108\\
 16.300& 2.344& 3.046& 16.249& 2.343& 3.032\\
 16.365& 2.260& 2.928& 16.326& 2.261& 2.918\\
 16.450& 2.155& 2.758& 16.427& 2.157& 2.752\\
 16.600& 2.108& 2.675& 16.597& 2.112& 2.674\\
 16.750& 2.082& 2.650& 16.759& 2.087& 2.652\\
 17.050& 2.076& 2.647& 17.066& 2.089& 2.651\\
 17.350& 2.094& 2.674& 17.359& 2.106& 2.676\\
 17.650& 2.130& 2.727& 17.650& 2.140& 2.727\\
 17.950& 2.176& 2.802& 17.949& 2.185& 2.802\\
 18.250& 2.245& 2.902& 18.247& 2.253& 2.901\\
 18.550& 2.324& 3.021& 18.546& 2.331& 3.020\\
 18.850& 2.405& 3.138& 18.838& 2.408& 3.135\\
 19.150& 2.516& 3.308& 19.103& 2.513& 3.296\\
 19.450& 2.627& 3.460& 19.383& 2.620& 3.442\\
 19.750& 2.733& 3.619& 19.707& 2.731& 3.608\\
 20.050& 2.870& 3.806& 19.965& 2.863& 3.783\\
 20.350& 2.995& 3.978& 20.264& 2.985& 3.954\\
 20.650& 3.120& 4.140& 20.568& 3.112& 4.118\\
 20.950& 3.248& 4.297& 20.862& 3.240& 4.274\\
 21.250& 3.341& 4.408& 21.161& 3.333& 4.384\\
 21.550& 3.446& 4.505& 21.459& 3.437& 4.480\\
 21.850& 3.564& 4.623& 21.758& 3.554& 4.598\\
 22.150& 3.651& 4.704& 22.056& 3.640& 4.679\\\hline
\end{tabular}
\tablecomments{$V-J$ and $V-K$ are the fiducial lines derived without any correction. For $(V-J)_{c}$ and $(V-K)_{c}$ a star-by-star correction has been applied appropriate to the atmospheric and chemical parameters of the individual stars and normalized to MSTO conditions ($T_{\mathrm{eff}}=6\,000\,\mbox{K}$, $\log g=4.5$).}
\end{table}

\begin{table}[H]
\centering
\caption{Empirical fiducial sequence for NGC6723.}
\begin{tabular}{cccccc}
\hline
\hline
$V$   &$(V-J)$	&   $(V-K)$ \\\hline
12.850& 2.878& 3.861\\
 13.150& 2.635& 3.593\\
 13.450& 2.473& 3.368\\
 13.750& 2.348& 3.170\\
 14.050& 2.271& 3.042\\
 14.350& 2.176& 2.904\\
 14.650& 2.061& 2.771\\
 14.950& 2.017& 2.684\\
 15.250& 1.959& 2.596\\
 15.550& 1.887& 2.503\\
 15.850& 1.836& 2.429\\
 16.150& 1.779& 2.365\\
 16.450& 1.757& 2.317\\
 16.750& 1.716& 2.264\\
 17.050& 1.691& 2.221\\
 17.350& 1.673& 2.195\\
 17.650& 1.636& 2.155\\
 17.950& 1.620& 2.112\\
 18.250& 1.567& 2.037\\
 18.400& 1.439& 1.866\\
 18.550& 1.335& 1.679\\
 18.850& 1.255& 1.592\\
 19.150& 1.249& 1.577\\
 19.450& 1.267& 1.591\\
 19.750& 1.305& 1.652\\
 20.050& 1.331& 1.714\\
 20.350& 1.419& 1.824\\
 20.650& 1.446& 1.931\\
 20.950& 1.529& 2.050\\
 21.250& 1.665& 2.248\\
 21.550& 1.831& 2.394\\
 21.850& 2.058& 2.695\\
 22.150& 2.220& 2.867\\
 22.450& 2.381& 3.053\\\hline
\end{tabular}
\end{table}

\begin{table}[H]
\centering
\caption{Empirical ZAHB fiducial sequence for M4.}
\begin{tabular}{cccccc}
\hline
\hline
$(V-J)$   &$(V-K)$	&   $V$ \\\hline
1.15&1.35& 13.78\\
 1.23&1.46& 13.58\\
 1.28&1.55& 13.52\\
 1.34&1.71& 13.49\\
 1.42&1.85& 13.50\\
 2.04&2.60& 13.46\\
 2.24&2.84& 13.47\\
 2.29&2.92& 13.47\\
 2.40&3.09& 13.40\\\hline
\end{tabular}
\end{table}

\begin{table}[H]
\centering
\caption{Empirical ZAHB fiducial sequence for NGC 6723.}
\begin{tabular}{cccccc}
\hline
\hline
$(V-J)$   &$(V-K)$	&   $V$ \\\hline
-0.04& -0.05& 16.81\\
 0.05& 0.05& 16.42\\
 0.18& 0.20& 15.88\\
 0.31& 0.30& 15.61\\
 0.41& 0.50& 15.53\\
 0.60& 0.76& 15.50\\
 1.13& 1.46& 15.53\\
 1.49& 1.87& 15.53\\
 1.62& 2.13& 15.33\\\hline
\end{tabular}
\end{table}

\section{Differential Extinction Tables}
\label{appendix_c}

\begin{table}[H]
\centering
\caption{Extinction coefficients for different temperatures.}
\begin{tabular}{lccccccccc}
\hline
\hline
$\log T_{\mathrm{eff}}\ \ $	&   ${A_{U}}\over{A_{V}}$	&   $A_{B}\over A_{V}$	&   $A_{R}\over A_{V}$	&   $A_{I}\over A_{V}$	&   $A_{J}\over A_{V}$	&   $A_{H}\over A_{V}$	&   $A_{K}\over A_{V}$	&  $A_{V}(abs)$	&  \\\hline
                           4.600& 1.578& 1.331& 0.832& 0.580& 0.284& 0.178& 0.115& 1.147 &\\
                          4.450& 1.581& 1.330&  0.831& 0.580& 0.283& 0.178& 0.115& 1.146\\
                          4.280& 1.572& 1.325&  0.828& 0.579& 0.284& 0.178& 0.115& 1.145\\
                          4.150& 1.564& 1.323&  0.828& 0.579& 0.284& 0.177& 0.115& 1.144\\
                          4.070& 1.559& 1.324&  0.829& 0.579& 0.286& 0.177& 0.116& 1.143\\
                          4.030& 1.555& 1.321&  0.829& 0.579& 0.284& 0.178& 0.115& 1.142\\
                          3.980& 1.552& 1.317&  0.828& 0.580& 0.284& 0.178& 0.115& 1.142\\
                         3.929& 1.550& 1.317&  0.826& 0.579& 0.284& 0.178& 0.115& 1.140 \\
                         3.858& 1.557& 1.315&  0.823& 0.581& 0.285& 0.179& 0.116& 1.137\\
                         3.815& 1.561& 1.314&  0.822& 0.580& 0.284& 0.179& 0.116& 1.135\\
                         3.778& 1.560& 1.311& 0.820& 0.580& 0.285& 0.180& 0.116& 1.134\\
                         3.764& 1.559& 1.310&  0.819& 0.580& 0.285& 0.180& 0.116& 1.133\\
                         3.747& 1.560& 1.306&  0.818& 0.581& 0.285& 0.180& 0.116& 1.132\\
                         3.715& 1.559& 1.305&  0.817& 0.582& 0.285& 0.180& 0.117& 1.129\\
                         3.689& 1.567& 1.300&  0.818& 0.582& 0.285& 0.180& 0.117& 1.125\\
                         3.653& 1.565& 1.303&  0.817& 0.584& 0.286& 0.181& 0.117& 1.122\\
                         3.638& 1.567& 1.298&  0.815& 0.584& 0.287& 0.181& 0.118& 1.120\\
                         3.622& 1.569& 1.299&  0.817& 0.586& 0.288& 0.182& 0.118& 1.116\\
                         3.602& 1.568& 1.299&  0.810& 0.585& 0.289& 0.182& 0.118& 1.114\\
                          3.580& 1.565& 1.298&  0.805& 0.583& 0.287& 0.182& 0.118& 1.115\\
                         3.566& 1.557& 1.296&  0.797& 0.582& 0.288& 0.181& 0.118& 1.116\\
                          3.550& 1.556& 1.299&  0.792& 0.581& 0.287& 0.182& 0.117& 1.116\\
                         3.519& 1.548& 1.295&  0.771& 0.576& 0.286& 0.181& 0.117& 1.119\\
                          3.470& 1.550& 1.285&  0.739& 0.571& 0.284& 0.181& 0.116& 1.120\\\hline
 \ \ \ \  -		&   1.664	&    1.321		&    0.819	&    0.594	& 0.276	& 0.176	& 0.112     	&	-	&	S\,98\\\hline
 \ \ \ \ -		&   1.569	&    1.323		&    0.751	&    0.478	& 0.282	& 0.192	& 0.116    		&	-	&	C\,89\\\hline
 \ \ \ \ -		&   -		&    1.32		&    0.82	&    0.59	& 0.29	& 0.23	& 0.11		&	-	&	B\,98\\\hline
 \ \ \ \ -		&   1.545	&    1.326		&    0.795	&    0.558	& 0.267    & 0.169 	& 0.114		&	-	&	M\,04\\\hline

\end{tabular}
\tablecomments{Relative extinctions for Johnson-Cousins $UBV(RI)_c$ (\citealt{Landolt92}) and 2MASS $J$, $H$, and $K_s$ filters for spectra with $R_V=3.10$, $\log g=4.5$, $\mbox{\hbox{\rm [Fe/H]}}=0.0$ and $E(B-V)=0.36$ for different temperatures. Since the ATLAS9 models cannot make accurate predictions for temperatures below 5\,000\,K, observed fluxes are used to calculate values in this table. S\,98: \citet{Schlegel98}; C\,89: \citet{Cardelli89}; B\,98: \citet{Bessell98}; M\,04: \citet{McCall04}.}
\label{APPX_a_relative_Teff}
\end{table}

\begin{table}[H]
\centering
\caption{Extinction coefficients for different metallicities.}
\begin{tabular}{lccccccccc}
\hline
\hline
$\hbox{\rm [Fe/H]}\ \ $	&   ${A_{U}}\over{A_{V}}$	&   $A_{B}\over A_{V}$	&   $A_{R}\over A_{V}$	&   $A_{I}\over A_{V}$	&   $A_{J}\over A_{V}$	&   $A_{H}\over A_{V}$	&   $A_{K}\over A_{V}$	&  $A_{V}(abs)$	&   \\\hline
$+0.5$& 1.565& 1.308& 0.821& 0.582& 0.284& 0.180& 0.116& 1.132& \\
       $+0.2$& 1.565& 1.310& 0.821& 0.582& 0.284& 0.180& 0.116& 1.132&\\
       $+0.0$& 1.565& 1.311& 0.821& 0.581& 0.284& 0.180& 0.116& 1.133&\\
       $-0.5$& 1.566& 1.313& 0.820& 0.581& 0.284& 0.180& 0.116& 1.133&\\
       $-1.0$& 1.567& 1.316& 0.820& 0.580& 0.284& 0.180& 0.116& 1.134&\\
       $-1.5$& 1.568& 1.318& 0.819& 0.580& 0.284& 0.179& 0.116& 1.134&\\
       $-2.0$& 1.569& 1.319& 0.819& 0.580& 0.284& 0.179& 0.116& 1.134&\\
       $-2.5$& 1.570& 1.320& 0.819& 0.580& 0.284& 0.179& 0.116& 1.134&\\
       $-4.0$& 1.571& 1.321& 0.819& 0.580& 0.284& 0.179& 0.116& 1.134&\\\hline    
 \ \ \ \  -		&   1.664	&    1.321		&    0.819	&    0.594	& 0.276	& 0.176	& 0.112     	&	-	&	S\,98\\\hline
 \ \ \ \ -		&   1.569	&    1.323		&    0.751	&    0.478	& 0.282	& 0.192	& 0.116    		&	-	&	C\,89\\\hline
 \ \ \ \ -		&   -		&    1.32		&    0.82	&    0.59	& 0.29	& 0.23	& 0.11		&	-	&	B\,98\\\hline
 \ \ \ \ -		&   1.545	&    1.326		&    0.795	&    0.558	& 0.267    & 0.169 	& 0.114		&	-	&	M\,04\\\hline
 \end{tabular}
\tablecomments{Relative extinctions for Johnson-Cousins $UBV(RI)_c$ (\citealt{Landolt92}) and 2MASS $J$, $H$, and $K_s$ filters for spectra with $T_{\mathrm{eff}}=6\,000\,\mbox{K}$, $\log g=4.5$, $R_V=3.10$ and $E(B-V)=0.36$ for different metallicities. S\,98: \citet{Schlegel98}; C\,89: \citet{Cardelli89}; B\,98: \citet{Bessell98}; M\,04: \citet{McCall04}.}
\label{APPX_a_relative_metals}    
\end{table}

\begin{table}[H]
\centering
\caption{Extinction coefficients for different dust type parameter $R_V$.}
\begin{tabular}{lccccccccc}
\hline%
\hline
$R_V \ \ $   &${A_{U}}\over{A_{V}}$	&   $A_{B}\over A_{V}$	&   $A_{R}\over A_{V}$	&   $A_{I}\over A_{V}$	&   $A_{J}\over A_{V}$	&   $A_{H}\over A_{V}$	&   $A_{K}\over A_{V}$	&  $A_{V}(abs)$	&   \\\hline
2.6& 1.687& 1.369&  0.803& 0.545& 0.260& 0.165& 0.106& 0.953& \\
 2.8& 1.633& 1.343&  0.811& 0.561& 0.271& 0.171& 0.110& 1.025\\
 3.0& 1.586& 1.321&  0.818& 0.575& 0.280& 0.177& 0.114& 1.097\\
 3.2& 1.545& 1.301&  0.824& 0.587& 0.288& 0.182& 0.118& 1.168\\
 3.4& 1.509& 1.284&  0.829& 0.598& 0.295& 0.187& 0.120& 1.240\\
 3.6& 1.477& 1.268&  0.834& 0.608& 0.301& 0.191& 0.123& 1.312\\
 3.8& 1.448& 1.255&  0.838& 0.616& 0.307& 0.194& 0.125& 1.384\\
 4.0& 1.421& 1.242&  0.842& 0.624& 0.312& 0.197& 0.127& 1.456\\
 4.2& 1.398& 1.231&  0.845& 0.631& 0.317& 0.200& 0.129& 1.527\\\hline
 \ \  -		&   1.664	&    1.321		&    0.819	&    0.594	& 0.276	& 0.176	& 0.112     	&	-	&	S\,98\\\hline
 \ \  -		&   1.569	&    1.323		&    0.751	&    0.478	& 0.282	& 0.192	& 0.116    		&	-	&	C\,89\\\hline
 \ \  -		&   -		&    1.32		&    0.82	&    0.59	& 0.29	& 0.23	& 0.11		&	-	&	B\,98\\\hline
 \ \  -		&   1.545	&    1.326		&    0.795	&    0.558	& 0.267    & 0.169 	& 0.114		&	-	&	M\,04\\\hline
\end{tabular}
\tablecomments{Relative extinctions for Johnson-Cousins $UBV(RI)_c$ (\citealt{Landolt92}) and 2MASS $J$, $H$, and $K_s$ filters for spectra with $T_{\mathrm{eff}}=6\,000\,\mbox{K}$, $\log g=4.5$, $\mbox{\hbox{\rm [Fe/H]}}=0.0$ and $E(B-V)=0.36$ for different $R_V$. S\,98: \citet{Schlegel98}; C\,89: \citet{Cardelli89}; B\,98: \citet{Bessell98}; M\,04: \citet{McCall04}.}
\label{APPX_a_relative_Rv}    
\end{table}%

\end{document}